\documentclass[journal=jacsat,manuscript=article]{achemso}

\usepackage[version=3]{mhchem} 
\usepackage{subcaption}
\usepackage{amsmath,amssymb}
\usepackage{filecontents}
\usepackage{natbib}

\author{Heather Mattie}
\email{hemattie@hsph.harvard.edu}
\affiliation{Biostatistics Department, Harvard T.H. Chan School of Public Health, 651 Huntington Ave., Boston, MA 02115}%

\author{Kenth Eng\o-Monsen}%
\affiliation{Telenor Research, D4d, Snar\o yveien 30, N-1360 Fornebu, Norway}%

\author{Rich Ling}
\affiliation{Wee Kim Wee School of Communication and Information, Nanyang Technological University, Singapore, 31 Nanyang Link, Singapore 637718}%

\author{Jukka-Pekka Onnela}
\affiliation{Biostatistics Department, Harvard T.H. Chan School of Public Health, 651 Huntington Ave., Boston, MA 02115}%

\title{The Social Bow Tie}

\begin{document}

\begin{abstract}
Understanding tie strength in social networks, and the factors that influence it, have received much attention in a myriad of disciplines for decades. Several models incorporating indicators of tie strength have been proposed and used to quantify relationships in social networks, and a standard set of structural network metrics have been applied to predominantly online social media sites to predict tie strength. Here, we introduce the concept of the ``social bow tie" framework, a small subgraph of the network that consists of a collection of nodes and ties that surround a tie of interest, forming a topological structure that resembles a bow tie. We also define several intuitive and interpretable metrics that quantify properties of the bow tie. We use random forests and regression models to predict categorical and continuous measures of tie strength from different properties of the bow tie, including nodal attributes. We also investigate what aspects of the bow tie are most predictive of tie strength in two distinct social networks: a collection of 75 rural villages in India and a nationwide call network of European mobile phone users. Our results indicate several of the bow tie metrics are highly predictive of tie strength, and we find the more the social circles of two individuals overlap, the stronger their tie, consistent with previous findings. However, we also find that the more tightly-knit their non-overlapping social circles, the weaker the tie. This new finding complements our current understanding of what drives the strength of ties in social networks.
\end{abstract}

\section{Introduction}
The strength of any kind of relationship between two individuals lies on a spectrum. People in general have a close relationship with only a few friends or family members, a somewhat weaker tie with a larger group of individuals with whom they interact less frequently, and an even weaker connection with a large number of casual acquaintances. This tradeoff between tie strength and the number of people a person is connected to through his or her ties was elegantly captured by Dunbar \cite{Dunbar}. Measuring and predicting tie strength, and moreover, understanding the factors that drive tie strength, has been an expanding area of interest, with increasing utility and complexity in the digital age, i.e., the ever-increasing forms of communication via mobile phones and social media. Knowledge of the strength of a tie, as well as the social dynamics contributing to tie strength, has been shown to increase the accuracy of link prediction, enhance the modeling of the spread of disease and information, and lead to more targeted marketing \cite{Li, Linyuan, Sa}. 

Several indicators of tie strength have been proposed, perhaps most notably by Mark Granovetter in his seminal work The Strength of Weak Ties \cite{Granovetter}. Granovetter differentiated between strong and weak ties and proposed the weak ties hypothesis: the stronger the tie between any two people, the higher the fraction of friends they have in common \cite{Granovetter}. Much of the current methodology centered on tie strength has stemmed from Granovetter's weak ties hypothesis and his proposed four dimensions of tie strength: the amount of time spent interacting with someone, the level of intimacy, the level of emotional intensity, and the level of reciprocity. More recently, three additional dimensions of tie strength have been proposed: 1) emotional support \cite{Marsden, Wellman}, 2) structural variables, i.e. network topology \cite{Ellison, Lin, Xiang}, and 3) social distance, i.e. the difference in socioeconomic status, education level, political affiliation, race, and gender \cite{He, Lin}. These categories have facilitated the definition and quantification of numerous possible predictors of tie strength; some generalizable to any network, and some specific to a limited number of social networks. Of importance to this analysis is a corresponding perspective outlined by Elizabeth Bott \cite{Bott} that suggests that the degree of clustering in an individual's network has the potential to draw them away from a dyadic tie if there are not mutual ties. 

Initially, highly generalizable similarity indices such as the number of common neighbors two nodes share, preferential attachment, and path distance were used to infer tie strength. These metrics were most commonly used for link prediction and were shown to provide some information regarding tie strength \cite{Linyuan, Papp}. However, it was quickly discovered that the addition of nodal attributes and other metrics not solely based on network topology greatly enhanced the measurement and prediction of tie strength \cite{Kahanda, Luarn}. Gilbert and Karahalios defined indicators of tie strength specific to a network of Facebook users and built a predictive model that achieved 85\% accuracy for binary tie strength (weak vs. strong) classification \cite{Gilbert}. They found that the act of communicating once leads to a significant increase in tie strength, and that educational difference plays a role in determining tie strength. Pappalardo et al. introduced a measure of tie strength using multiple online social networks and found that the strength of a tie is related to the number of interactions between the two individuals \cite{Papp}. In addition, several studies have shown that frequent communication, both online and offline, is positively related to tie strength \cite{Marsden, Wiese}.

While previous studies have provided advances and valuable insights, they suffer from a binary definition of tie strength (weak vs strong), low diversity in the types of social networks studied (the vast majority being social media sites), and non-representative samples. In this work, we propose a decomposition of a social network into an ensemble of interconnected ``social bow ties,'' constellations consisting of nodes and ties that surround each network tie. We call any such subgraph a ``social bow tie'' because the topological structure that surrounds each tie resembles a bow tie. We also introduce several simple metrics that quantify properties of the bow tie. Further, we use random forests and linear regression to build models that predict categorical and continuous measures of tie strength from different properties of the bow tie, including nodal attributes (covariates) of the nodes included in the bow tie. We apply our framework to two social networks, a collection of 75 social networks from the villages of Karnataka, India, and a call network of European mobile phone subscribers. We find that the bow tie framework contributes to more accurate predictions of tie strength and provides insights on which metrics are the most informative of tie strength. Specifically, we find that the larger the proportion of shared friends, the stronger the tie, and the more clustered the individual friendship circles (consisting of non-overlapping friends), the weaker the tie. Consequently, these findings provide evidence to support both the weak ties hypothesis and a generalized version of the Bott hypothesis \cite{Bott}.


\section{Methods}
\subsection{Data Description}
We analyzed two social network data sets. The first data set is social network data collected in 2006 from 75 villages located in 5 districts in rural southern Karnataka, India. The data were collected through household and individual surveys as part of a study by Banerjee et al. \cite{Banerjee}. Of relevance for this study, the survey included social network data along 12 dimensions: friends or relatives who visit the respondent's home, friends or relatives the respondent visits, any kin in the village, non-relatives with whom the respondent socializes, those from whom who the respondent receives medical advice, with whom who the respondent goes to temple to pray, from whom the respondent would borrow money, to whom the respondent would lend money, from whom the respondent would borrow material goods from, to whom the respondent would lend material goods, from whom the respondent gets advice, and to whom the respondent gives advice. It is worth noting that these forms of interaction are largely face-to-face, unlike the mediated material from the call detail records (CDRs) described below. Additionally, a proportion of villagers were given individual surveys that recorded age and sex, among other attributes. 

For this data set, we define the strength of a tie as the number of distinct types of social relationships reported to exist between the two individuals. For example, if individual $i$ borrows money from individual $j$ and in addition gives advice to individual $j$, the weight of the (undirected) tie between $i$ and $j$ would be equal to 2. If $i$ and $j$ also attend temple together, their tie strength would be 3 and so on, with a minimum strength of 1 and a maximum strength of 12 for any tie. Note that a tie strength of 0 implies that the two individuals are not connected by any kind of social tie. We denote the strength of a tie between individuals $i$ and $j$ as $w_{ij}$. Because we ignore the directionality of ties, our definition of tie strength is symmetric. 

The second data set consists of call detail records (CDRs) from a mobile phone provider in an undisclosed European country where 68\% of citizens own a smartphone and 85\% own a cellular phone. The data examined here span a period of three months in 2013, and each record consists of the following daily aggregate communication summaries for pairs of individuals: the date, anonymized caller ID, anonymized callee ID, daily call duration (in minutes), daily number of calls, daily number of text messages (SMS), and daily number of multimedia messages (MMS). Age, sex, and billing zip codes were available for a large majority (72.3\%) of individuals. 

An undirected, weighted call network was created from the records by first summing the call durations between any two individuals over the three-month period. If two individuals spoke on the phone at least once during the period, we connected them with an edge of strength $w_{ij}$, where the value of edge strength was set to the total amount of time spent on the phone with one another. Since tie strength is defined in terms of absolute time, it does not take into account the total amount of time each individual spends on the phone, which makes it somewhat difficult to quantify the relative strength of ties since the strength of a tie is not measured on the same scale either for individuals or pairs of individuals. We therefore normalized tie strength and represent it with two measurements: one that represents tie strength from the perspective of individual $i$, and one that represents tie strength from the perspective of individual $j$. Specifically, for each tie, the first measurement of tie strength is the total call duration ($w_{ij}$) divided by the total time individual $i$ spends on the phone $s_i$, the strength of node $i$. similarly, the second measurement of tie strength is the total call duration divided by the total time individual $j$ spends on the phone $s_j$, the strength of node $j$. Dividing total call duration by the strength of each focal node results in a consistent definition of tie strength. We denote these new tie strength measurements as $y_{ij}$ and $y_{ji}$. We created another summary measure of tie strength by taking the average of $y_{ij}$ and $y_{ji}$, and we denote this $z_{ij} =(y_{ij} + y_{ji})/2$.

\subsection{Bow Tie Framework}
\begin{figure}[t]
\centering
\includegraphics[scale=0.35]{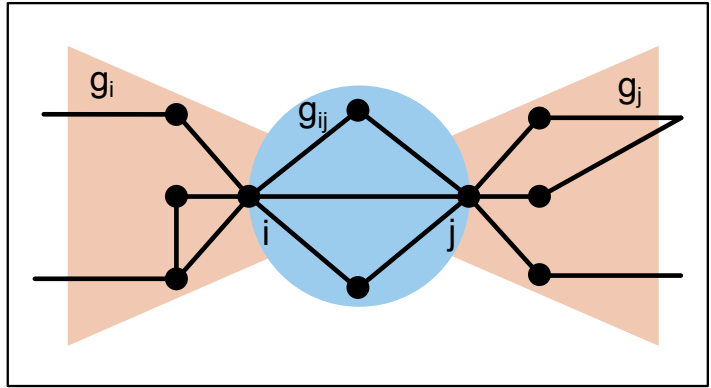}
\caption{A simple example of the social bow tie $G_{ij}$. The blue circle contains the nodes and edges that comprise the overlapping friendship circle of the focal nodes $i$ and $j$, denoted $g_{ij}$. The parts of the bow tie shaded in orange contain the individual (non-overlapping) social circles of the focal nodes, denoted $g_i$ for node $i$ and $g_j$ for node $j$.}
\label{fig:bowtie}
\end{figure}

To introduce the ``bow tie'' structure, consider a weighted social network $G$, which may be directed or undirected, and consider a tie with weight $w_{ij}$ that connects two individuals $i$ and $j$. We call these two individuals the \emph{focal nodes} of the bow tie. We use the term \emph{focal tie} to refer to the tie that links them. We start by partitioning $i$'s friends and $j$'s friends into three disjoint sets. Group $i$, denoted $g_i$, contains the nodes that are connected to only $i$; group $j$, denoted $g_j$, contains nodes that are connected to only $j$; and group $ij$, denoted $g_{ij}$, contains nodes that are connected to both $i$ and $j$. These three groups jointly make up the shared and non-shared friends of $i$ and $j$. We call this structure the $ij$ \emph{bow tie}. Formally, the groups $g_i$, $g_j$ and $g_{ij}$ are induced subgraphs, where the node sets that induce them are the neighbors of $i$, the neighbors of $j$, and the common neighbors of $i$ and $j$, respectively. The bow tie $ij$, denoted by $G_{ij}$, is the subgraph that is induced by the union of all neighbors of $i$ and $j$. Note that $G_{ij}$ is more than the sum of $g_i$, $g_j$ and $g_{ij}$: in addition to containing the same set of nodes and ties as those subgraphs do, it also contains the inter-group ties among this set of nodes, i.e., the ties linking nodes across $g_i$, $g_j$ and $g_{ij}$. Important to our analysis below is the hierarchical structure of the bow tie: at the upper level of hierarchy we have the bow tie $G_{ij}$; at the intermediate level, we have the three groups, $g_i$, $g_j$ and $g_{ij}$; and at the lowest level we have the nodes and ties from which each group is composed. A simple example of the bow tie structure surrounding nodes $i$ and $j$ is shown in Figure \ref{fig:bowtie}. 

The localized nature of the bow tie framework gives rise to several topological metrics that can be used to predict tie strength and find evidence for or against both the weak ties hypothesis and the Bott hypothesis. We include unweighted \cite{Onnela} and weighted \cite{Mattie1} edge overlap, which we denote $o_{ij}$ and $\tilde o_{ij}$, respectively. Unweighted overlap is defined as in \eqref{eq:overlap}, and weighted overlap as in \eqref{eq:woverlap}.
 
\begin{equation}\label{eq:overlap}
o_{ij} = \frac{n_{ij}}{k_i + k_j - 2 - n_{ij}}
\end{equation}

\begin{equation}\label{eq:woverlap}
\tilde{o}_{ij} = \frac{\sum^{n_{ij}}_{k=1}(w_{ik} + w_{jk})}{s_i + s_j - 2w_{ij}}
\end{equation}

Here, $n_{ij}$ is the number of common (shared) friends of nodes $i$ and $j$, $k_i$ $(k_j)$ denotes the degree, or number of connections, node $i$ $(j)$ has, $w_{ij}$ denotes the weight associated with the tie between nodes $i$ and $j$, and $s_i$ $(s_j)$ denotes the strength of node $i$ $(j)$. In accordance with the weak ties hypothesis, we expect both $o_{ij}$ and $\tilde{o}_{ij}$ to be positively associated with tie strength, i.e., that tie strength $w_{ij}$, increases as the number of shared friends increases. Metrics based on customized versions of the clustering coefficients of $i$ and $j$ are used, where the calculation of a clustering coefficient is limited to the non-shared friends of each node, i.e., for node $i$, the nodes and edges in $g_i$ are used to calculate the clustering coefficient of $i$, and similarly, $g_j$ is used for node $j$. We denote the sum and absolute difference of these quantities as $cc^S_{ij}$ and $cc^D_{ij}$ for the unweighted clustering coefficients, and $\tilde cc^S_{ij}$ and $\tilde cc^D_{ij}$ for the weighted clustering coefficients. Here, we use the definition of weighted clustering coefficient provided by Saramäki et. al. \cite{Saramaki}. Specifically, the weights of ties are considered and the metric reflects how large triangle weights are compared to a network maximum. Other predictors include the sum and absolute difference in the degrees of $i$ and $j$ ($k^S_{ij}$ and $k^D_{ij}$), the sum and absolute difference in the strengths of $i$ and $j$ ($s^S_{ij}$ and $s^D_{ij}$), the number of nodes and edges in $g_{ij}$  ($n_{ij}$ and $e_{ij}$), and the sum and absolute difference in the number of nodes and the number of edges in $g_i$ and $g_j$ ($n^S_{ij}$, $n^D_{ij}$, $e^S_{ij}$ and $e^D_{ij}$). With these definitions, we can represent Bott's hypothesis in two different ways; using $s^S_{ij}$ and $cc^S_{ij}$. Bott suggests that the more close-knit the non-overlapping social circles of two connected individuals, the weaker the tie between them. Translating this to our setting, we expect tie strength to be negatively associated with $s^S_{ij}$ and $cc^S_{ij}$. Specifically, as the clustering and strength of ties among individuals in $g_i$ and $g_j$ increases, tie strength ($w_{ij}$) decreases. Finally, predictors created from the attributes of $i$ and $j$ include the sum and absolute difference in the ages of $i$ and $j$ ($a^S_{ij}$ and $a^D_{ij}$), the paired sex category (male-male, female-female, female-male) denoted $\textrm{I}_{\textrm{MM}}$, $\textrm{I}_{\textrm{FF}}$ and $\textrm{I}_{\textrm{FM}}$ respectively, and an indicator if $i$ and $j$ have the same billing zip code, denoted $Z_{ij}$. See Table \ref{table:preds} for a detailed description of each variable.


\begin{table}[t]
\caption{Descriptions of tie strength predictors.}\label{table:preds}
\centering
\begin{tabular}{|l|l|}
\hline
Predictor & Description\\
\hline
\hline
$k^S_{ij}$  & Sum of the degrees of $i$ and $j$ ($k_i + k_j$)\\
$k^D_{ij}$  & Absolute difference in the degrees of $i$ and $j$ ($|k_i - k_j|$)\\
$s^S_{ij}$  & Sum of the strengths of $i$ and $j$ ($s_i + s_j$)\\
$s^D_{ij}$  & Absolute difference in the strengths of $i$ and $j$ ($|s_i - s_j|$)\\
$cc^S_{ij}$  & Sum of the clustering coefficients of $i$ and $j$\\
$cc^D_{ij}$ & Absolute difference in the clustering coefficients of $i$ and $j$\\
$\tilde cc^S_{ij}$  & Sum of the weighted clustering coefficients of $i$ and $j$\\
$\tilde cc^D_{ij}$ & Absolute difference in the weighted clustering coefficients of $i$ and $j$\\
$a^S_{ij}$  & Sum of the ages of $i$ and $j$\\
$a^D_{ij}$ & Absolute difference in the ages of $i$ and $j$\\
$\textrm{Sex}_{ij}$ & Categorical variable indicating a male-male, female-female, or female-male tie\\
$\textrm{I}_{\textrm{MM}}$ & Indicator variable of a male-male tie \\ 
$\textrm{I}_{\textrm{FF}}$ & Indicator variable of a female-female tie\\
$\textrm{I}_{\textrm{FM}}$ & Indicator variable of a female-male tie\\
$Z_{ij}$ & Indicator if i and j have the same billing zip code \\
$o_{ij}$& Unweighted overlap of edge between $i$ and $j$\\
$\tilde{o}_{ij}$ & Weighted overlap of edge between $i$ and $j$\\
$n_{ij}$ & Number of common friends of $i$ and $j$ \\
$e_{ij}$ & Number of edges among the common friends of $i$ and $j$\\
$n^S_{ij}$ & Sum of the number of nodes in $g_i$ and $g_j$\\
$n^D_{ij}$ & Absolute difference in the number of nodes in $g_i$ and $g_j$\\
$e^S_{ij}$ & Sum of the number of edges in $g_i$ and $g_j$\\
$e^D_{ij}$ & Absolute difference in the number of edges in $g_i$ and $g_j$\\
\hline
\end{tabular}
\end{table}

To predict tie strength and study how it is associated with different metrics, we used regression as well as Random Forest (RF) regression and classification \cite{Breiman}. For the India social network, tie strength is discrete with $w_{ij} \in \{1, \dots, 12\}$. Thus, the weight of a tie can be viewed as a categorical outcome, allowing RF classification and Poisson regression to be used to predict tie strength, or as continuous with RF regression used for prediction. For the CDR call network, tie strength is most naturally treated as a continuous variable, and we used RF regression and linear regression to predict both measures of tie strength. 

In addition to ordinary least squares (OLS) regression, least absolute shrinkage and selection operator (LASSO) and ridge regression were used to fit more parsimonious and interpretable models as well as increase prediction accuracy. Before using LASSO and ridge regression, all data was centered around the mean and 10-fold cross validation was performed to select the best tuning parameters; denoted $\lambda^L$ for LASSO and $\lambda^R$ for ridge regression. For RF classification, the number of trees used was 200, and the maximum number of features (covariates) considered when splitting a node was $\sqrt{n}$ where $n$ is the total number of features. For RF regression, 200 trees were used and the maximum number of features considered when splitting a node was $n$.

Nodal attributes were expected to be informative of tie strength and were therefore included in the models. Each individual's attributes were known for the subset of the India data set used for analysis. However, individuals in the CDR call network could have any combination of age, sex and billing zip code information missing. We used RF classification to impute sex and RF regression to impute age. Because of the abundance of billing zip code possibilities, rather than imputing billing zip code directly, we created a paired billing zip code dichotomous variable equal to 1 if the two focal nodes had the same billing zip code and 0 if they did not. We then used RF classification to impute paired billing zip code.

\section{Results}
\subsection{India Social Network}
The India network contained 69,444 nodes, of which 16,984 (24.5\%) had full attribute information available, and 294,778 edges after the removal of isolated ties. Of these, 37,714 (12.8\%) edges were between two individuals with complete attribute information available, and comprised our sample of edges for analysis. We discovered tie strength had a bimodal distribution with $\approx$46\% of ties having a strength of 12. This was due to the fact that the majority (96\%) of ties between individuals living in the same household had a weight of 12. We decided to exclude ties between individuals from the same household and only included cross-household ties. This resulted in a Poisson distribution of tie strength and a total of 21,945 ties.

\begin{figure}
\begin{subfigure}{.5\textwidth}
  \centering
  \includegraphics[width=.75\linewidth]{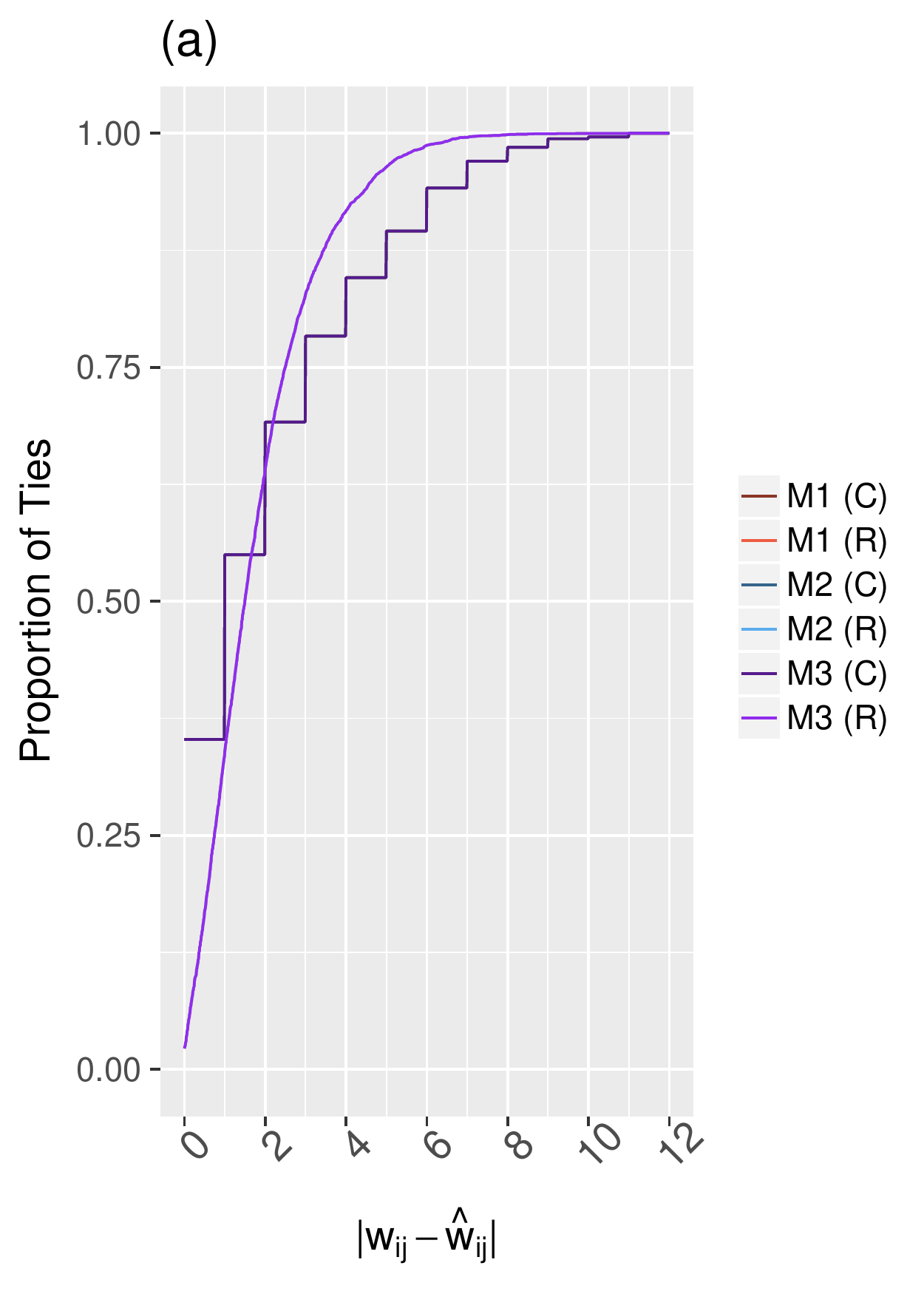}
  \caption*{}
  \label{fig:ind_all}
\end{subfigure}%
\begin{subfigure}{.5\textwidth}
  \centering
  \includegraphics[width=.75\linewidth]{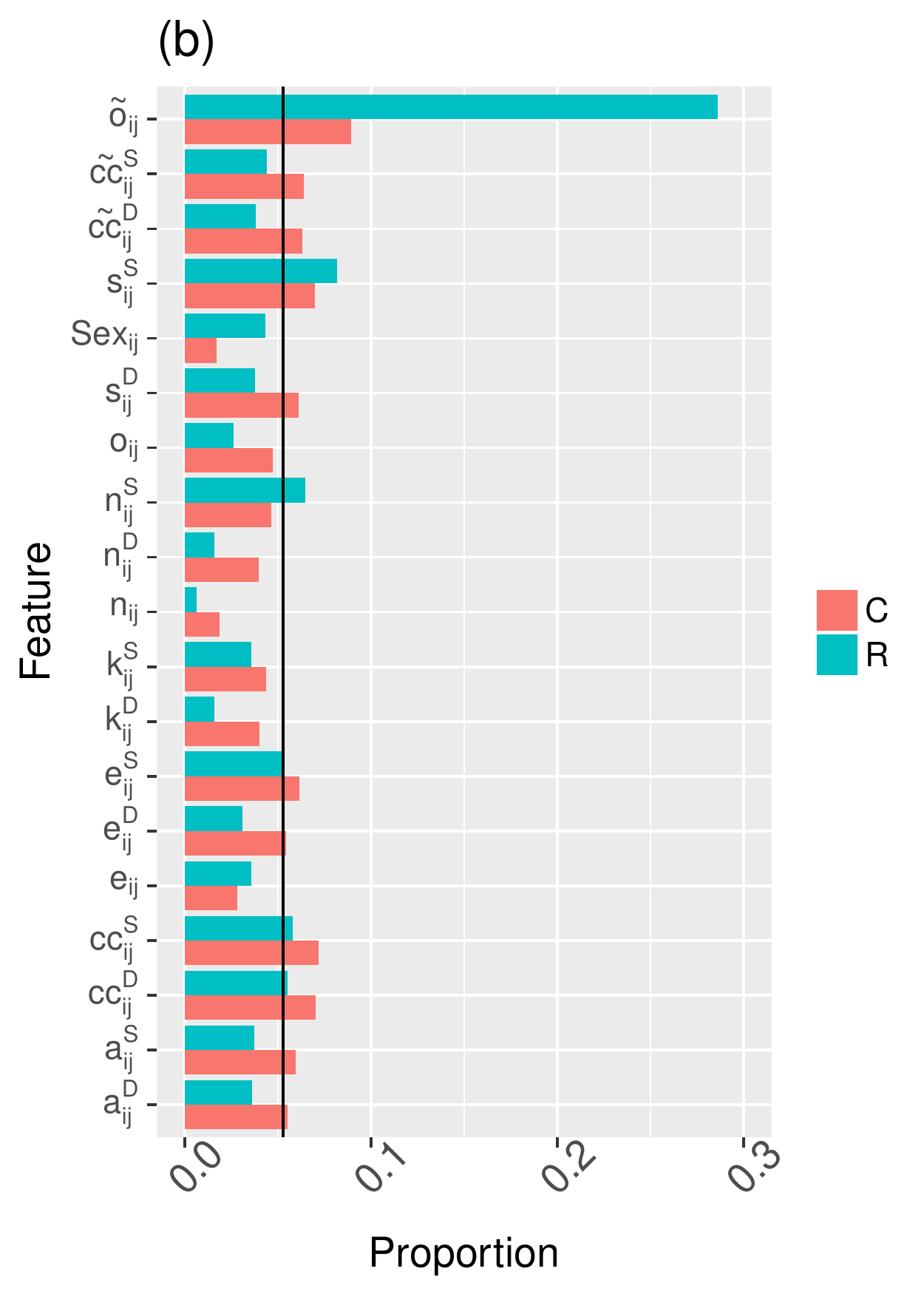}
  \caption*{}
  \label{fig:ind_fi1}
\end{subfigure}%
\\
\begin{subfigure}{.5\textwidth}
  \centering
  \includegraphics[width=.75\linewidth]{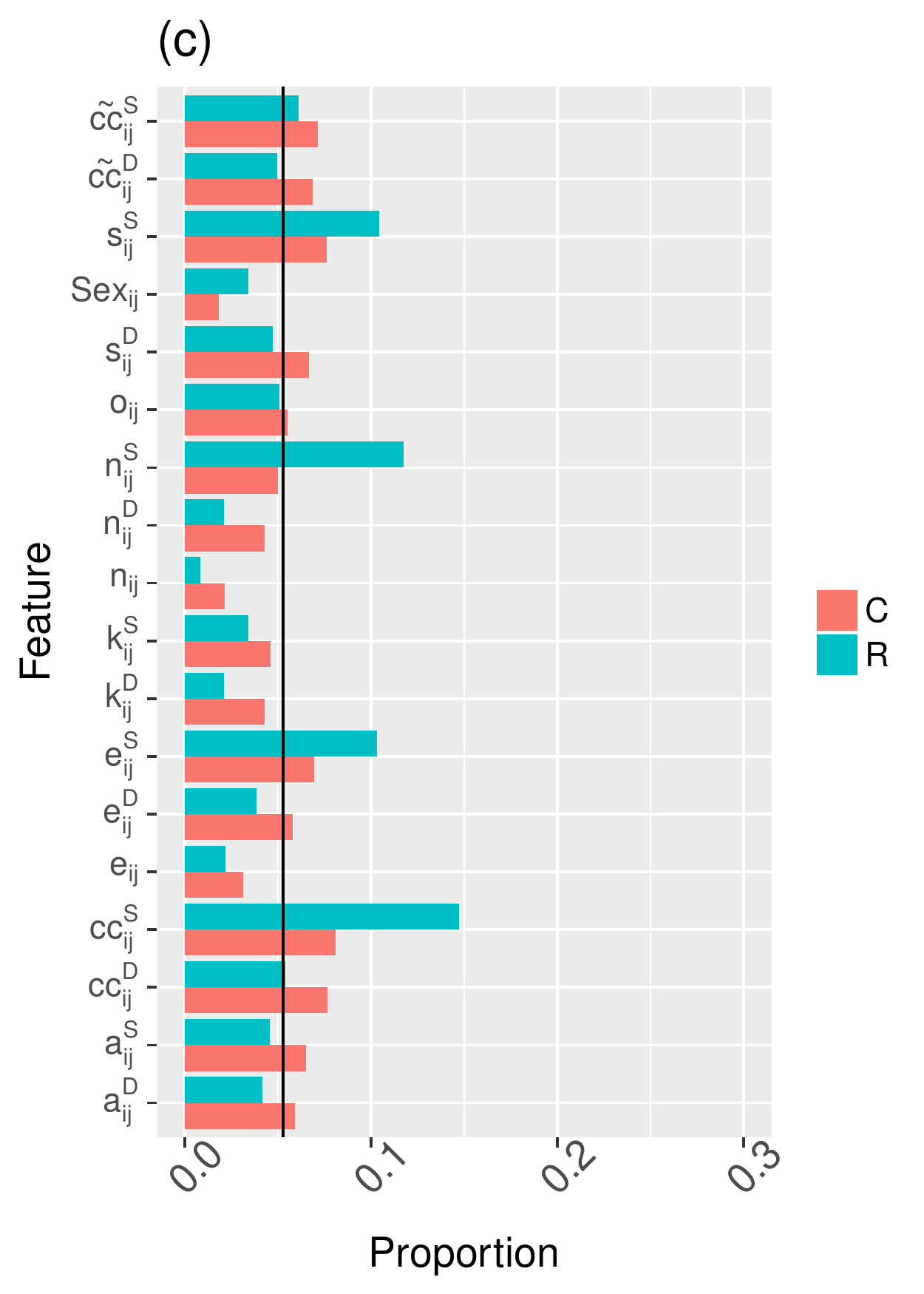}
  \caption*{}
  \label{fig:ind_fi2}
\end{subfigure}%
\begin{subfigure}{.5\textwidth}
  \centering
  \includegraphics[width=.75\linewidth]{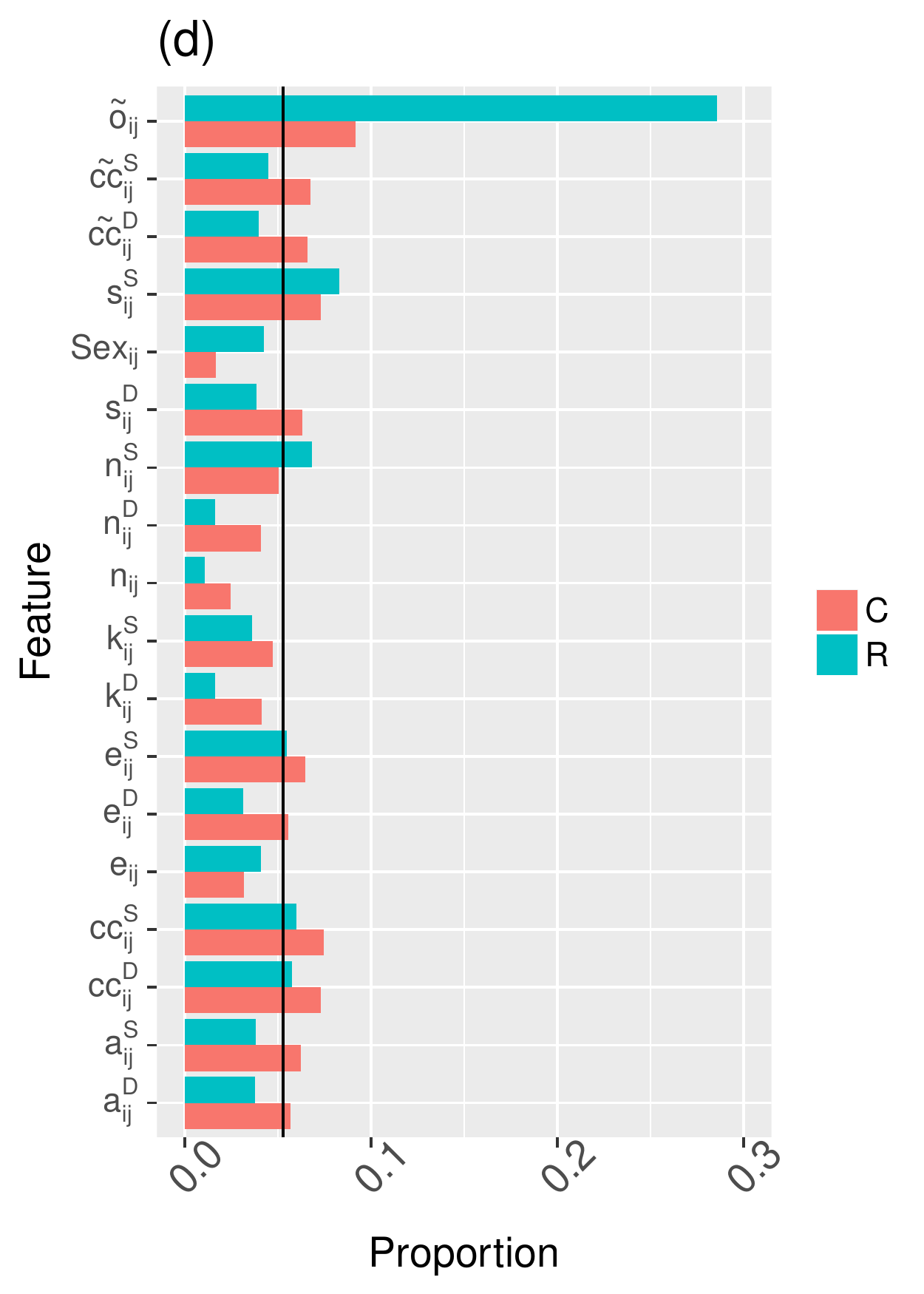}
  \caption*{}
  \label{fig:ind_fi3}
\end{subfigure}
\caption{Accuracy and feature importance plots for the India social network. Accuracy, measured as the absolute difference between empirical tie strength ($w_{ij}$) and predicted tie strength ($\hat w_{ij}$), for Models 1-3 using both RF regression (R) and classification (C) after imputation is shown in (a). Feature importance using RF regression and classification after imputation are shown for Model 1 (b), Model 2 (c) and Model 3 (d). The horizontal bars represent how informative the predictor is with a longer bar meaning more informative. The black vertical line represents the value of an equilibrium or null importance if every predictor were equally informative.}
\label{fig:india_all}
\end{figure}

RF regression and classification were used to fit three models both before and after nodal attribute imputation, where ties with complete attribute information available were included in the analysis before imputation and all ties were included after imputation. Model 1 is the full model and includes all covariates described in Table \ref{table:preds} with the exception of $Z_{ij}$ since it is specific to the CDR data set; Model 2 includes all covariates except weighted overlap; and Model 3 includes all covariates except unweighted overlap. It has been shown that categorical predictors do not need to be split into multiple dichotomous covariates (referred to as dummy variables) when implementing RF if there are a small number of them and their cardinality is low \cite{Breiman, Hastie}. Therefore, the variable $\textrm{Sex}$ was not split into two separate dummy variables due to its low cardinality and it being the single categorical predictor. Accuracy was measured as the residual, the absolute difference between empirical tie strength ($w_{ij}$) and predicted tie strength ($\hat w_{ij}$). Figure \ref{fig:india_all}(a) shows the accuracy of RF regression and classification for all models. Note that only two lines are visible, one for RF regression and one for RF classification since the accuracy of all models is indistinguishable. Within one unit of tie strength, an accuracy of 36.4\% and 55.3\% was achieved by RF regression and classification, respectively. 

Feature importance for each of the three models for both RF regression and classification is shown in Figure \ref{fig:india_all}(b)-(d). The horizontal bars represent how informative the predictor is with a longer bar meaning more informative. The black vertical line represents the value of an equilibrium or null importance if every predictor were equally informative. For both classification and regression, weighted overlap ($\tilde o_{ij}$) is the most informative variable in models 1 and 3, and the sum of the clustering coefficients ($cc^S_{ij}$) is the most informative in model 2, followed by the sum of the number of friends in the non-overlapping social circles ($n^S_{ij}$). These results provide evidence that the proposed indicators of tie strength in the Weak Ties and Bott hypotheses (the overlap of friendship circles and the amount of clustering in the non-overlapping friendship circles) are predictive of tie strength. 

Poisson regression was used to model the associations between tie strength and each of the predictors, and the coefficients of significant predictors with magnitude greater than 0.2 are reported in \eqref{eq:india_eq}. The predictors with the largest magnitudes include $\tilde o_{ij}$, $cc^S_{ij}$, and $\textrm{I}_{\textrm{FM}}$. Weighted overlap is positively associated with tie strength, illustrating the greater the proportion of strength among overlapping friends of the focal nodes, the stronger the tie between the focal nodes, and showing evidence to support Granovetter's hypothesis. The sum of the clustering coefficients of the focal nodes is positively associated with tie strength, meaning tie strength decreases as the amount of clustering in the non-overlapping friendship circles increases. This provides quantitative evidence of Bott's hypothesis in a novel population. Finally, the predictor $\textrm{I}_{\textrm{FM}}$ is negatively associated with tie strength, indicating that on average, female-male ties are weaker than male-male ties, which were used a reference group.

\begin{equation}\label{eq:india_eq}
log(\mathop{\mathbb{E}}[w_{ij}]) = 1.62 + 2.41\tilde o_{ij} -1.38cc^S_{ij} -0.2\text{I}_{\text{FM}}
\end{equation}

\subsection{CDR Call Network}
The CDR call network contained 2,276,495 nodes and 12,345,848 edges. Age was available for 89.25\% of the individuals and had a mean of 48.2 (sd = 18.2) years. Of the 89.03\% of individuals whose sex was recorded, 52.51\% were male. Billing zip code was available for 99.35\% of individuals. Due to the large size of the network, a random sample of 500,000 edges was drawn. After the removal of isolated ties, a total of 496,941 edges remained. Full attribute information was available for both focal nodes for 359,367 (72.3\%) edges. 

\begin{figure}
\begin{subfigure}{.5\textwidth}
  \centering
  \includegraphics[width=.75\linewidth]{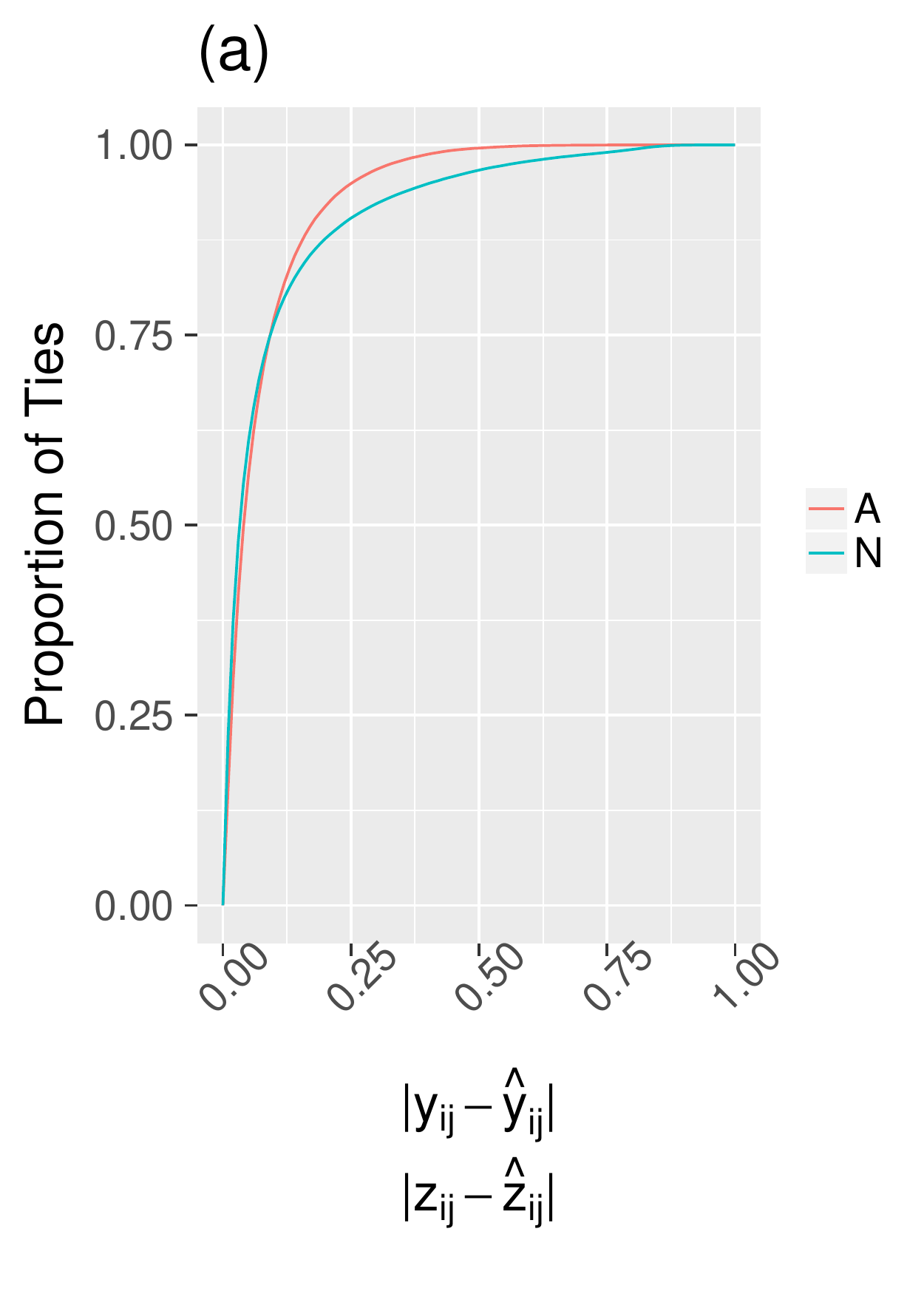}
  \caption*{}
  \label{fig:cdr_all}
\end{subfigure}%
\begin{subfigure}{.5\textwidth}
  \centering
  \includegraphics[width=.75\linewidth]{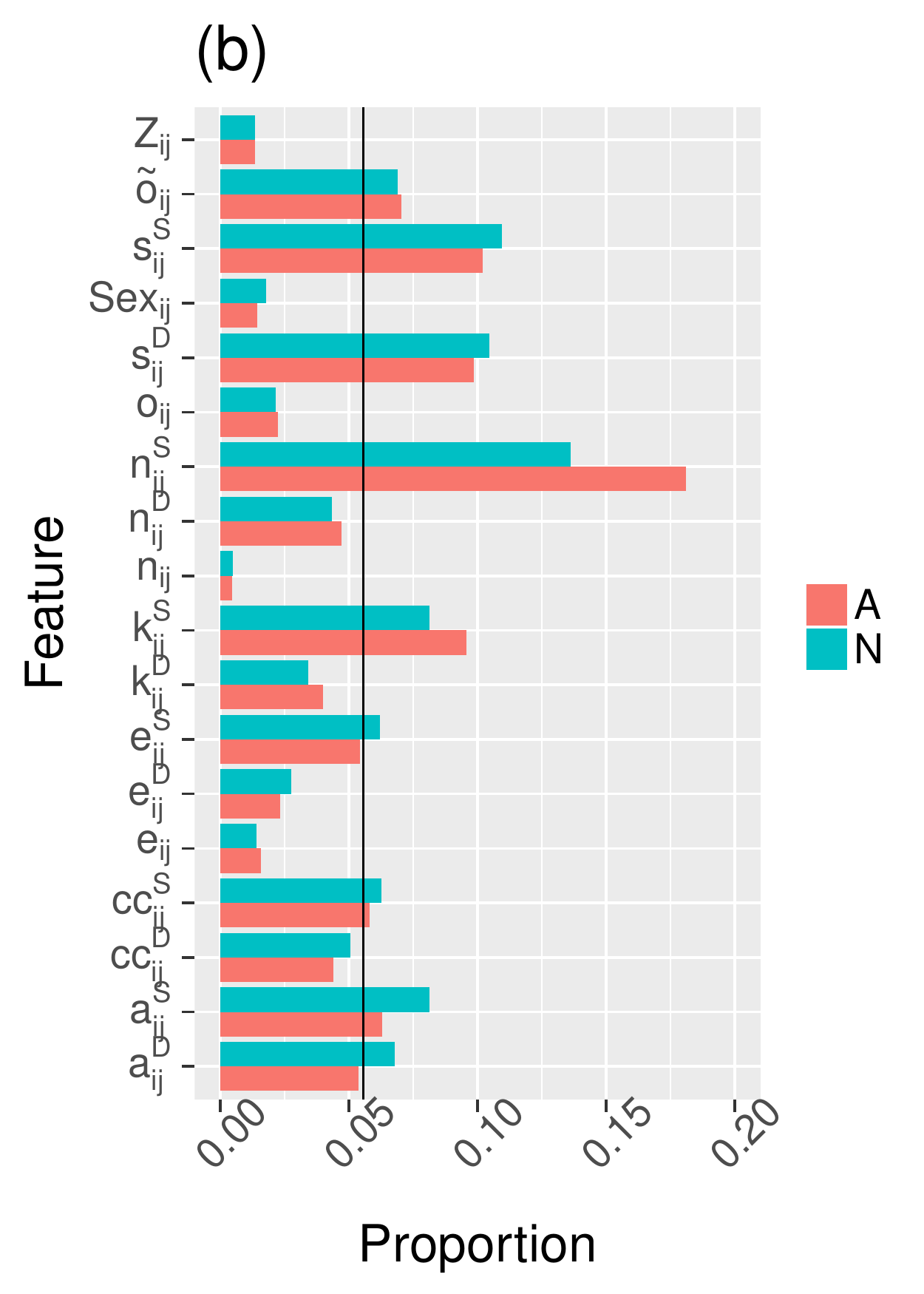}
  \caption*{}
  \label{fig:cdr_fi1}
\end{subfigure}%
\\
\begin{subfigure}{.5\textwidth}
  \centering
  \includegraphics[width=.75\linewidth]{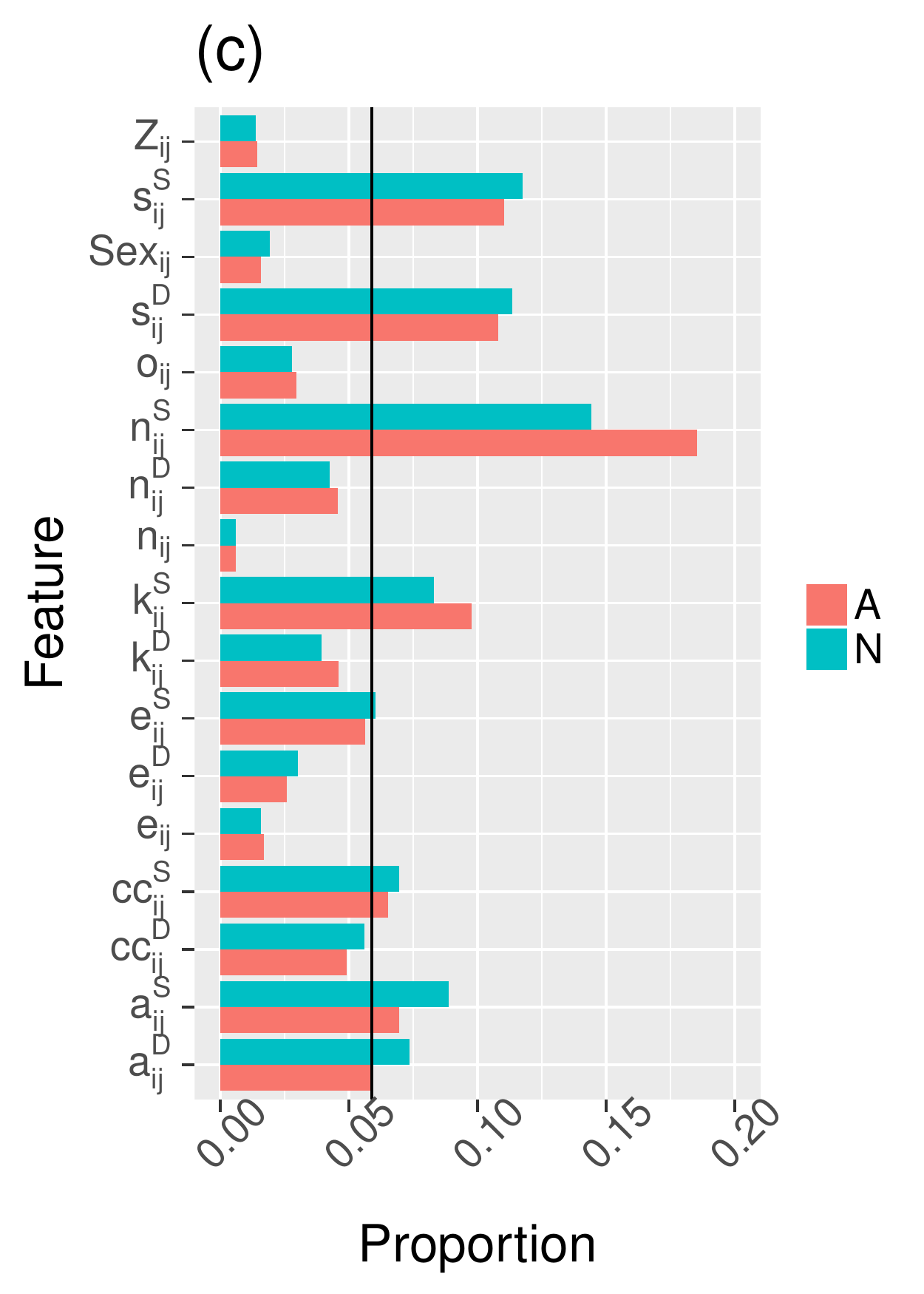}
  \caption*{}
  \label{fig:cdr_fi2}
\end{subfigure}%
\begin{subfigure}{.5\textwidth}
  \centering
  \includegraphics[width=.75\linewidth]{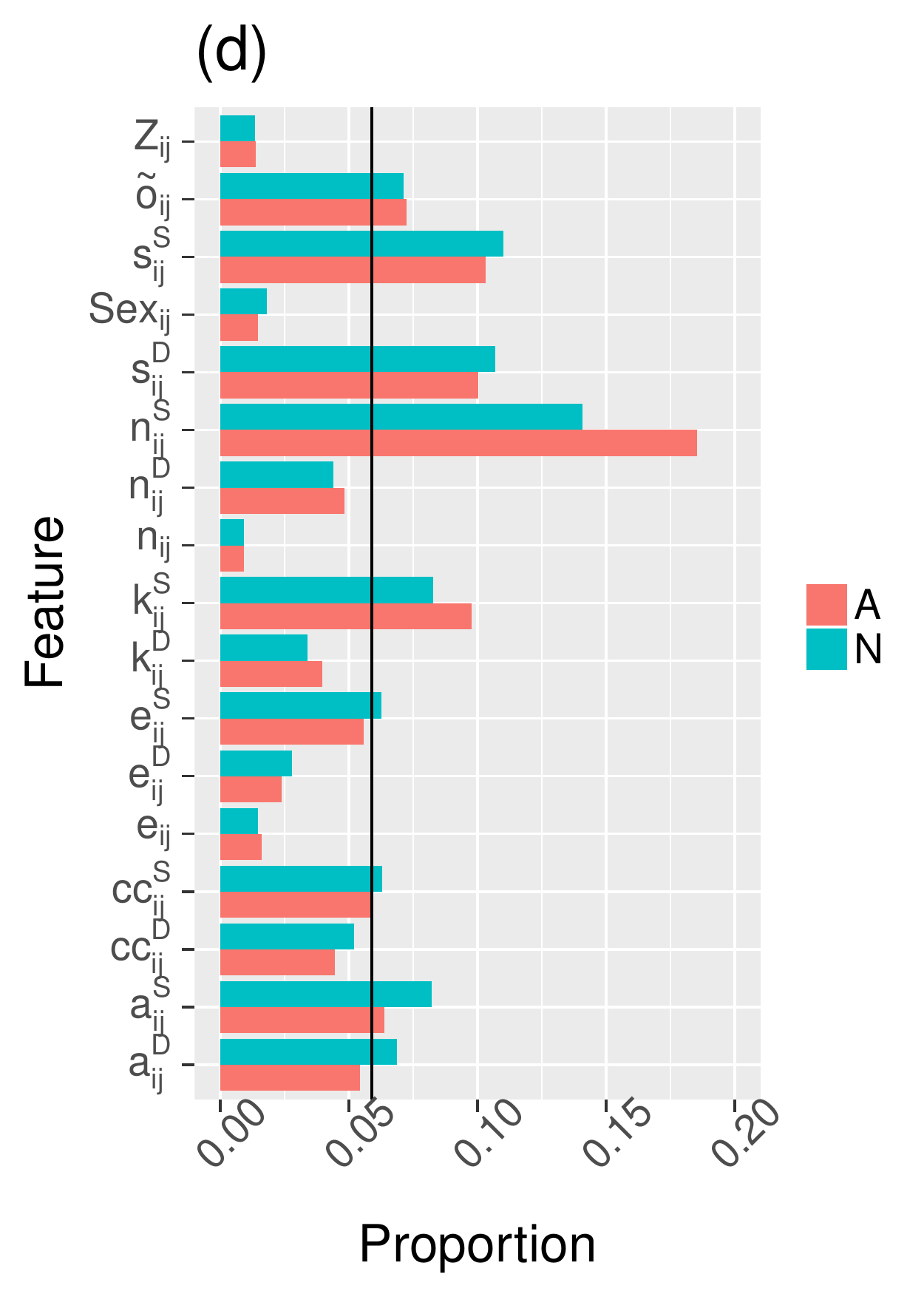}
  \caption*{}
  \label{fig:cdr_fi3}
\end{subfigure}
\caption{Accuracy and feature importance plots for the CDR call network with normalized (N) and averaged (A) tie strengths. Accuracy, measured as the absolute difference between empirical tie strength ($y_{ij}, z_{ij}$) and predicted tie strength ($\hat y_{ij}, \hat z_{ij}$), for all three models using RF regression after imputation is shown in (a). Note that only one curve is visible for each strength measure since the accuracy of all three models is indistinguishable. Feature importance using RF regression after imputation are shown for Model 1 (b), Model 2 (c) and Model 3 (d).}
\label{fig:CDR_all}
\end{figure}

Similar to the India data set, three models were fit with RF regression both before and after nodal attribute imputation for each measure of tie strength and are denoted Models 1-3. Figure \ref{fig:CDR_all}(a) shows the accuracy for RF regression after imputation for all three models and each measure of tie strength. The difference in accuracy for all models is very minimal and only one curve is visible for each tie strength measure. Within 0.05 units (a 5\% difference between empirical and predicted tie strength), an accuracy of 61\% was achieved for normalized tie strength, and 56.7\% for averaged tie strength. Within 0.1 units, an accuracy of 76.5\% was achieved for normalized tie strength and 77.3\% for averaged tie strength. Accuracy for all models and both tie strength measurements before and after imputation are shown in Figures S1(a) and S2(a) in the supporting information (SI). Imputation has a smaller impact on accuracy for this data set in all cases. 

Feature importance for each of the three models after imputation is shown in Figure \ref{fig:CDR_all}(b)-(d). The black vertical line represents the value of importance if every predictor were equally informative. The most informative predictors in each model are $s^S_{ij}$, $s^D_{ij}$, $n^S_{ij}$ and $k^S_{ij}$, with $\tilde o_{ij}$ and $a^S_{ij}$ slightly more informative than the null importance value in models 1 and 3. This suggests focal node strength, degree and number of non-overlapping friends are the aspects of the bow tie most predictive of tie strength in this network. Feature importance plots for all models and all tie strength measures before and after imputation are presented in Figures S1(b)-(d) and S2(b)-(d) in the SI.

For each measure of tie strength, three different models, denoted Models A - C, were fit using linear regression methods following imputation. Model A denotes the full model that was fit using OLS regression. Model B was fit using LASSO and Model C using ridge regression. Because the distributions of normalized and averaged tie strength are highly skewed for this data set, we first log-transformed each measure of tie strength and then centered them around the mean. All predictors were standardized (centered around the mean with unit variance) before fitting models B and C. Implementing LASSO and ridge regression require the selection of tuning parameters that determine the extent of shrinkage administered when calculating coefficient estimates. As the tuning parameter approaches 0, the corresponding coefficient estimates match the OLS estimates. In this extreme, the amount of bias is minimal, if nonexistent, but the amount of variance is comparatively high. As the tuning parameter is increased, the values of the coefficients decrease and approach 0 once the tuning parameter is sufficiently large. In this extreme, bias is increased but variance in the estimates is decreased. The optimal choice for a tuning parameter balances the amount of bias and variance and can be selected via cross-validation. We performed 10-fold cross validation to select values of the tuning parameters $\lambda^L$ and $\lambda^R$. The values of the LASSO coefficients as a function of $\lambda^L$ and, as a more interpretable measure, the $l_1$ penalty $\| \hat \beta_L \| / \| \hat \beta \|_1$ which represents the amount of shrinkage, are shown in Figures S3(a)-(b) and S4(a)-(b) in the SI. The values of the ridge regression coefficients as a function of $\lambda^R$ and the $l_2$ penalty $\| \hat \beta_R \| / \| \hat \beta \|_2$ are shown in Figures S3(c)-(d) and S4(c)-(d) in the SI. Significant predictors, their coefficients, adjusted $R^2$ values and the values of the tuning parameters for models B and C are presented in Table S1 in the SI. \eqref{eqn:OLS1}, \eqref{eqn:LASSO1} and \eqref{eqn:RIDGE1} show the fitted regression equations for normalized tie strength, $y_{ij}$, for OLS, LASSO and ridge regression respectively. Similarly, \eqref{eqn:OLS2}, \eqref{eqn:LASSO2} and \eqref{eqn:RIDGE2} show the fitted regression equations for averaged tie strength, $z_{ij}$, for OLS, LASSO and ridge regression respectively.

\begin{eqnarray}
\mathop{\mathbb{E}}(y_{ij})_{OLS} &=&  - 0.35k^D_{ij} - 0.25s^S_{ij} + 0.29cc^D_{ij} +0.23Z_{ij}+ 0.27o_{ij}\label{eqn:OLS1}\\
\mathop{\mathbb{E}}(y_{ij})_{LASSO} &=&  - 0.33k^D_{ij} - 0.25s^S_{ij} + 0.23cc^D_{ij} +0.23Z_{ij} + 0.21o_{ij} \label{eqn:LASSO1}\\
\mathop{\mathbb{E}}(y_{ij})_{RIDGE} &=&  - 0.35k^D_{ij} - 0.25s^S_{ij} + 0.29cc^D_{ij} +0.23Z_{ij}+0.27o_{ij} \label{eqn:RIDGE1}\\
\mathop{\mathbb{E}}(z_{ij})_{OLS} &=& - 0.35k^D_{ij} - 0.25s^S_{ij} + 0.29cc^D_{ij} +0.23Z_{ij} +0.27o_{ij} -0.2s^D_{ij} \label{eqn:OLS2}\\
\mathop{\mathbb{E}}(z_{ij})_{LASSO} &=& - 0.21k^D_{ij} - 0.39s^S_{ij} + 0.24cc^D_{ij} +0.23Z_{ij} \label{eqn:LASSO2}\\
\mathop{\mathbb{E}}(z_{ij})_{RIDGE} &= &- 0.27k^D_{ij} - 0.49s^S_{ij} + 0.36cc^D_{ij} +0.24Z_{ij} +0.28o_{ij} +0.31s^D_{ij} \label{eqn:RIDGE2}
\end{eqnarray}

For normalized tie strength, $\lambda^R$ was sufficiently large such that no shrinkage was implemented, and the estimated ridge regression coefficients are equivalent to the OLS estimates. The amount of LASSO shrinkage was approximately 12\%, resulting in slightly different coefficient estimates. In all models, $o_{ij}$, $k^D_{ij}$, $s^S_{ij}$, $cc^D_{ij}$ and $Z_{ij}$ were significantly associated with tie strength. Edge overlap is positively associated with tie strength in all models, showing that as the proportion of common friends two individuals share increases, so does the strength of the tie between the two individuals, supporting Granovetter's hypothesis. Tie strength is negatively associated with $s^S_{ij}$ which suggests that as the focal nodes expand their social circles and the time spent interacting with friends, the weaker the tie between them; more evidence to support Bott's hypothesis. The positive association between $Z_{ij}$ and tie strength implies having the same billing zip code increases the strength of a tie and could suggest a geographical impact on tie strength.

Here, $cc^D_{ij}$ is positively associated with tie strength meaning the more dissimilar the non-overlapping clustering coefficients of the focal nodes, the stronger their tie. Lastly, the $R^2$ values for these models are on the lower side (0.112 on average). This could be due to the network being constructed with phone-based communication rather than face-to-face interactions among highly clustered villagers. Furthermore, quantifying tie strength for CDR data is currently still rather ambiguous; the operationalization of using communication as a proxy for tie strength has not yet been validated \cite{Wiese}. An alternate measure of tie strength may increase the $R^2$ values.

\section{Discussion}
In this work, we introduce the social bow tie; a novel framework we use to perform a comprehensive analysis of the association between network structure and tie strength. Our framework decomposes a social network into a collection of nodes and ties immediately surrounding each network tie. This utilization of local structure produces easily interpretable metrics that quantify social perspectives of tie strength and allows for analyses that are computationally feasible for networks of any size. Through machine learning and regression methods including LASSO and ridge regression, we determine which properties of the bow tie structure are the most predictive of tie strength in two different types of social networks; a contact network of Indian villagers and a nationwide call network of European mobile phone users.

Overall, both data sets provide evidence to support the weak ties hypothesis and the Bott hypothesis. Following Granovetter, we find that the more friends two individuals share, the stronger their tie. Following Bott, the more tightly-knit their individual social circles, the weaker their tie. In addition, we find that the bow tie framework provides metrics that predict tie strength with high accuracy for both networks. 

In future work, it would be interesting to apply the bow tie framework to a social network of married couples. In this case the dominant strong tie has properties that are not seen in more casual social ties, namely the individuals constitute a particularly strongly defined social institution that has both emotional (romantic attachment) as well as structural (e.g. common responsibility for children and common ownership of capital investments such as a home) elements that provide it resiliency. This would enable testing of the original version of Bott's hypothesis, rather than a generalized form as we present here. It would also be interesting to test if the strength of in-person ties behaves similarly for the mobile phone call network. 

\begin{acknowledgement}

The authors thank Telenor for making available anonymized data for this study, and Banerjee et. al. for making the India data set publicly available.

\end{acknowledgement}

\section{Supporting Information}
Here we present additional accuracy and feature importance plots for tie strength in the CDR call network, as well as figures demonstrating the choice of shrinkage parameters $\lambda^L$ and $\lambda^R$ for both measures of tie strength using 10-fold cross-validation. Finally, a table of values for $\lambda^L$, $\lambda^R$ and all regression coefficients is presented.

\begin{figure}[b]
\begin{subfigure}{.5\textwidth}
  \centering
  \includegraphics[width=0.7\linewidth]{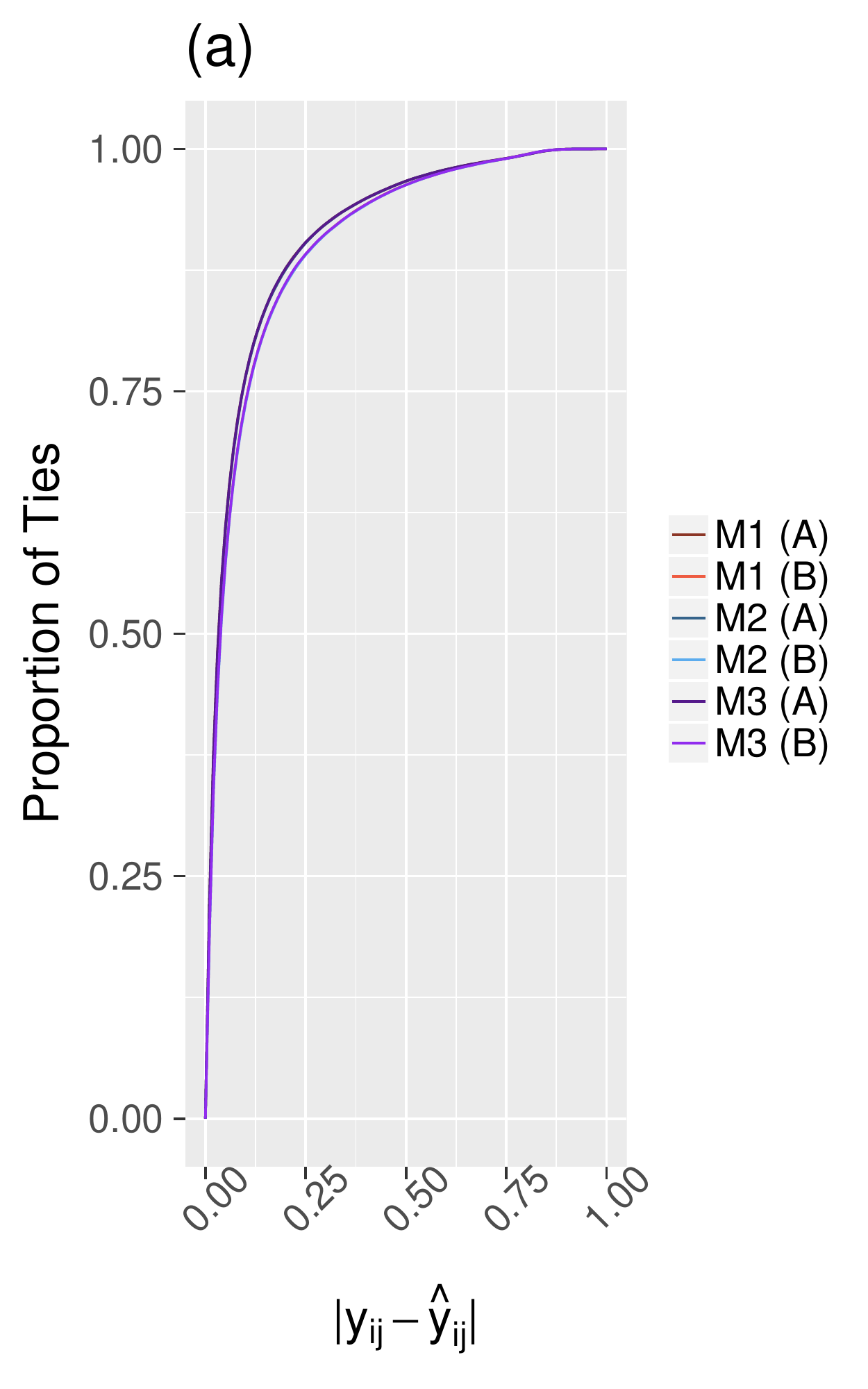}
  \caption*{}
\end{subfigure}%
\begin{subfigure}{.5\textwidth}
  \centering
  \includegraphics[width=0.7\linewidth]{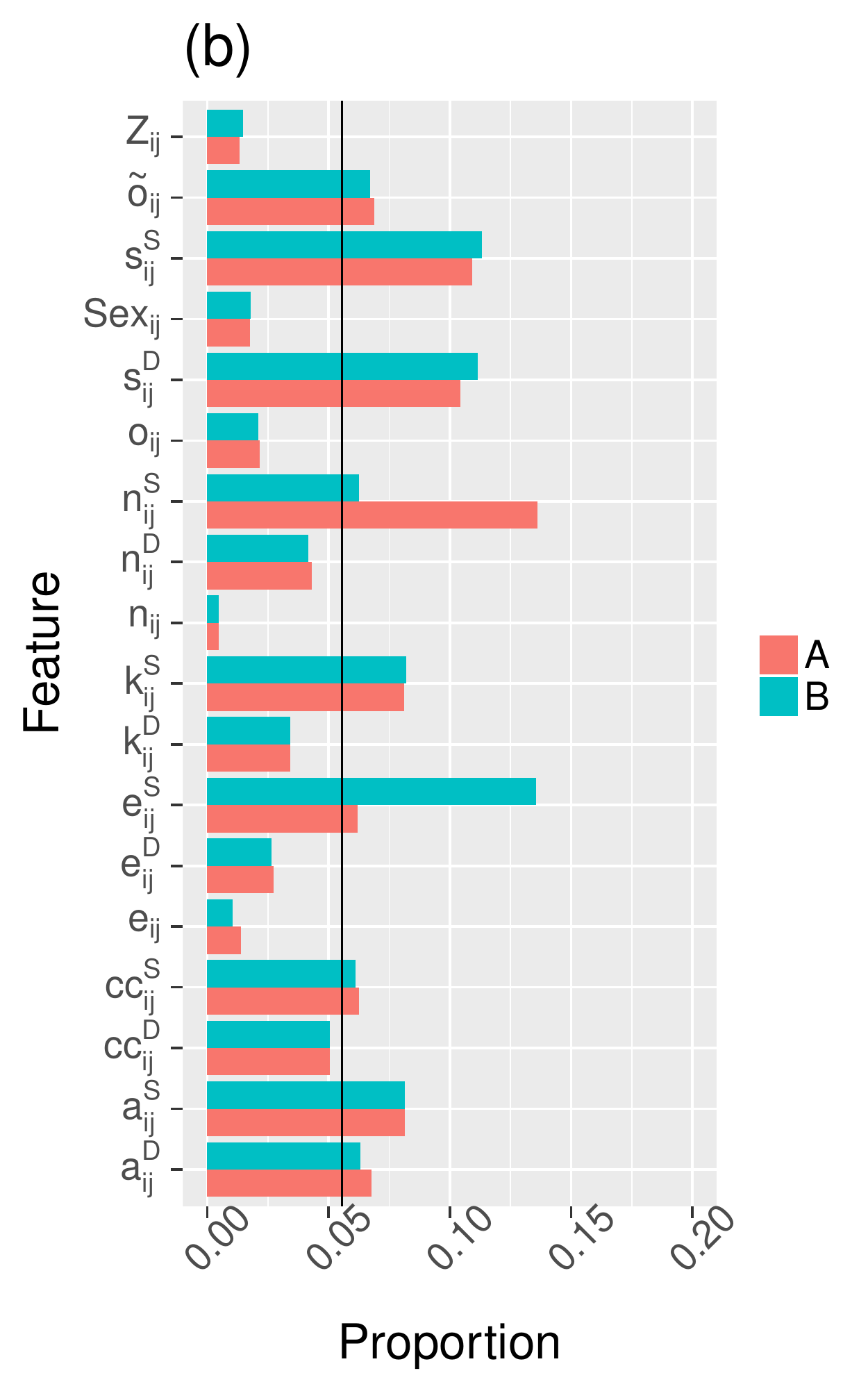}
  \caption*{}
\end{subfigure}%
\\
\begin{subfigure}{.5\textwidth}
  \centering
  \includegraphics[width=0.7\linewidth]{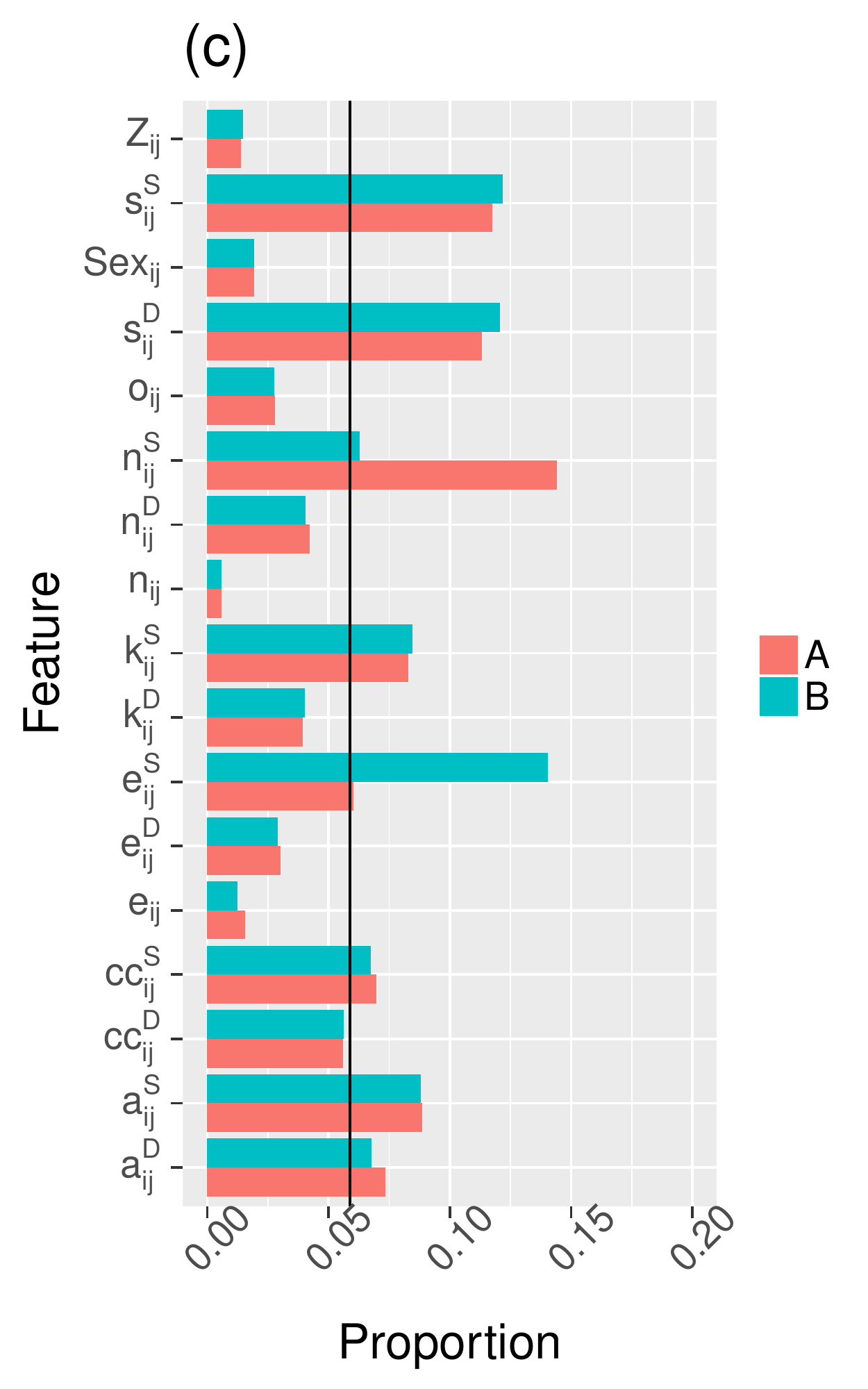}
  \caption*{}
\end{subfigure}%
\begin{subfigure}{.5\textwidth}
  \centering
  \includegraphics[width=0.7\linewidth]{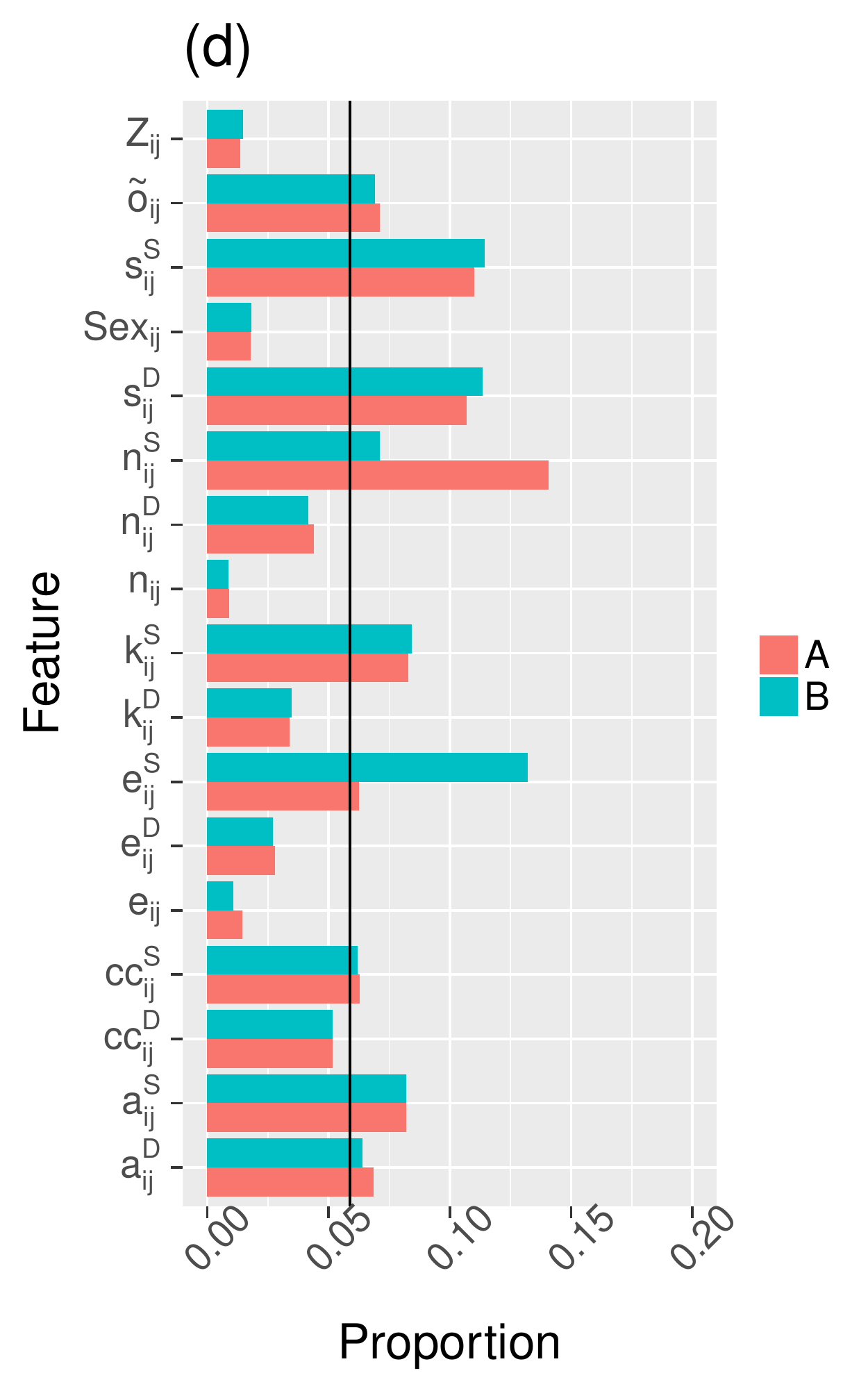}
  \caption*{}
\end{subfigure}
\caption*{\textbf{Fig. S1.} Accuracy and feature importance plots for the normalized tie strength ($y_{ij}$) CDR call network. Accuracy using RF regression before (B) and after (A) imputation for all three models is shown in (a). Feature importance using RF regression before and after imputation are shown for Model 1 (b), Model 2 (c) and Model 3 (d). 
}
\end{figure}

\begin{figure}
\begin{subfigure}{.5\textwidth}
  \centering
  \includegraphics[width=0.7\linewidth]{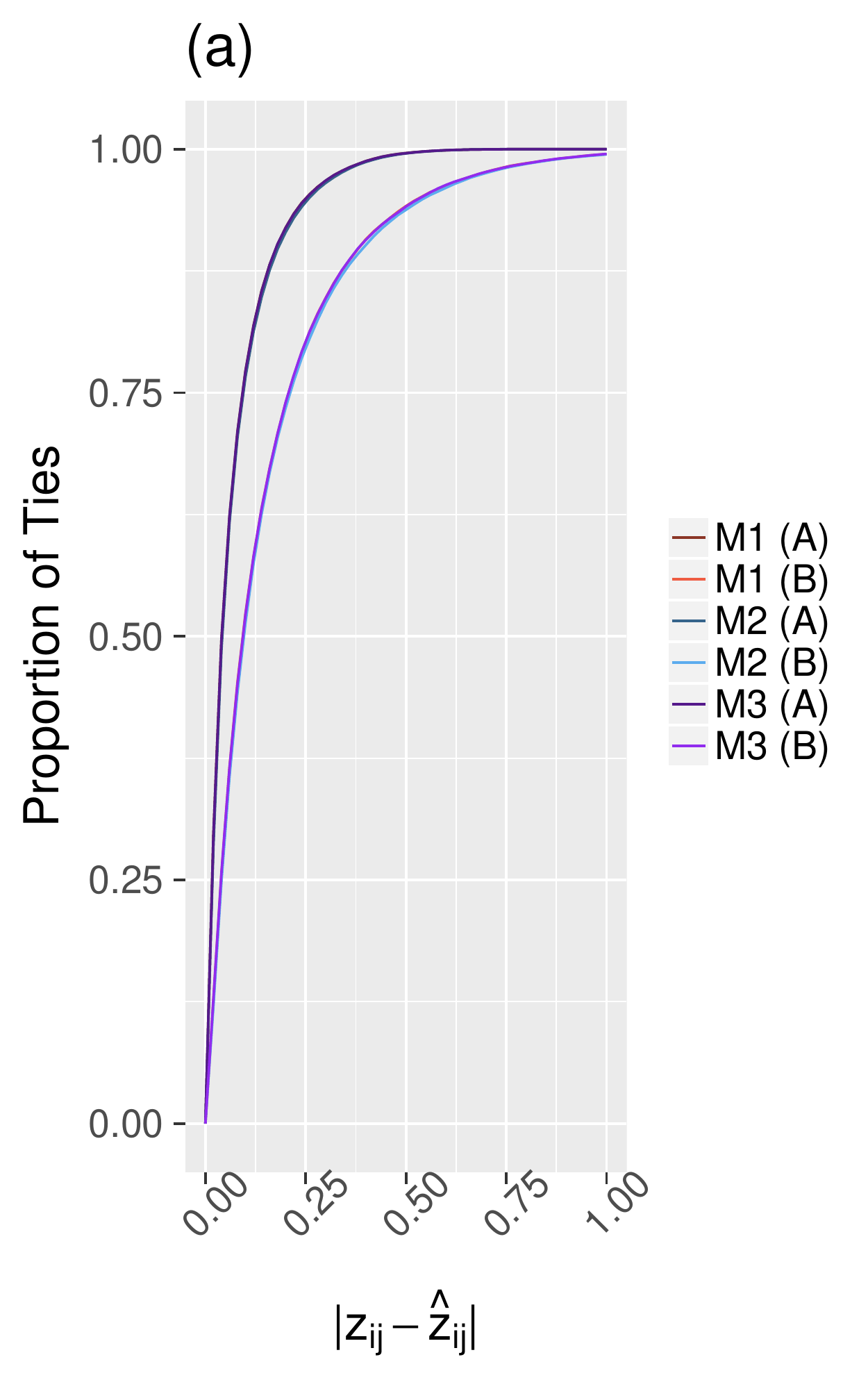}
  \caption*{}
\end{subfigure}%
\begin{subfigure}{.5\textwidth}
  \centering
  \includegraphics[width=0.7\linewidth]{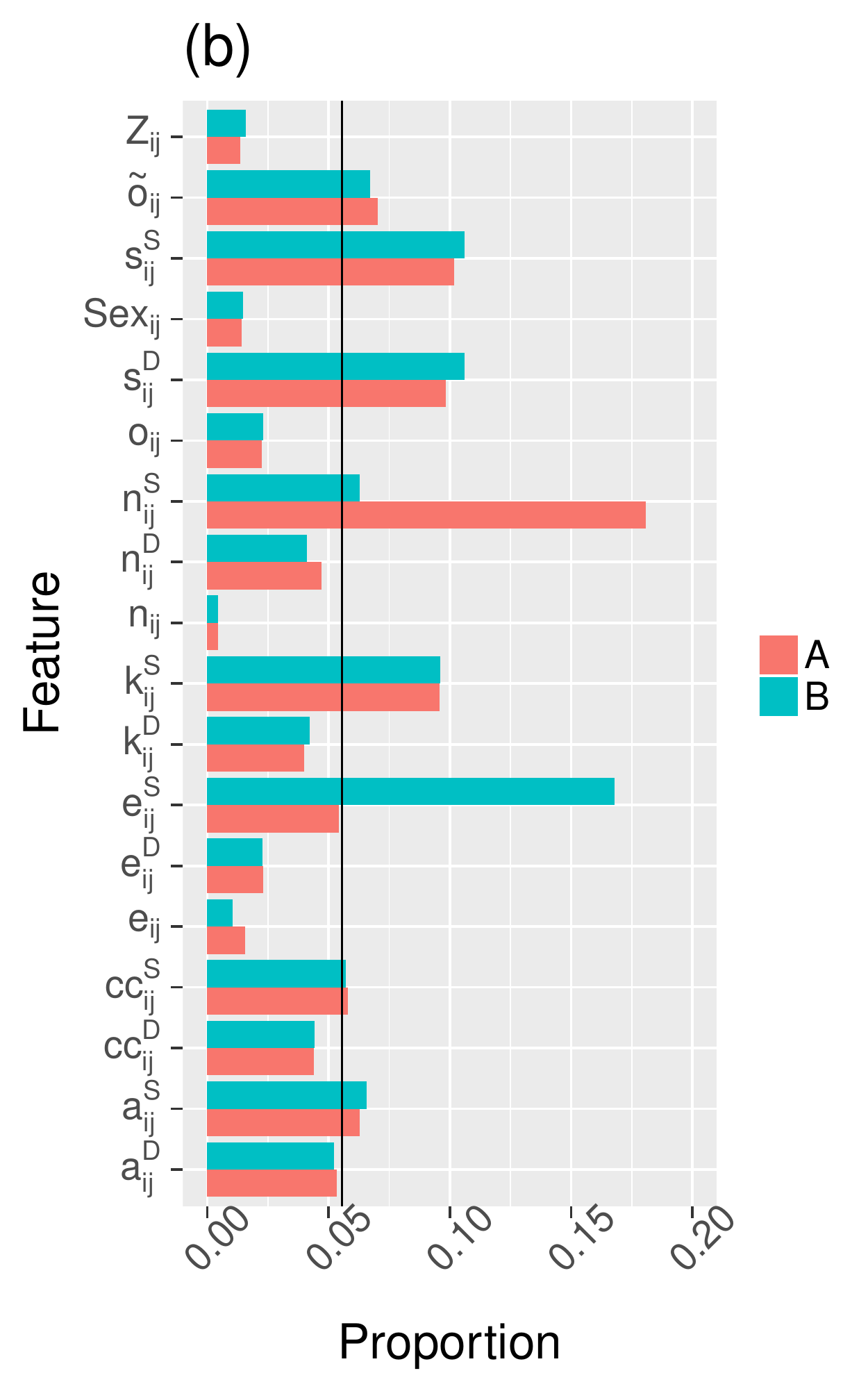}
  \caption*{}
\end{subfigure}%
\\
\begin{subfigure}{.5\textwidth}
  \centering
  \includegraphics[width=0.7\linewidth]{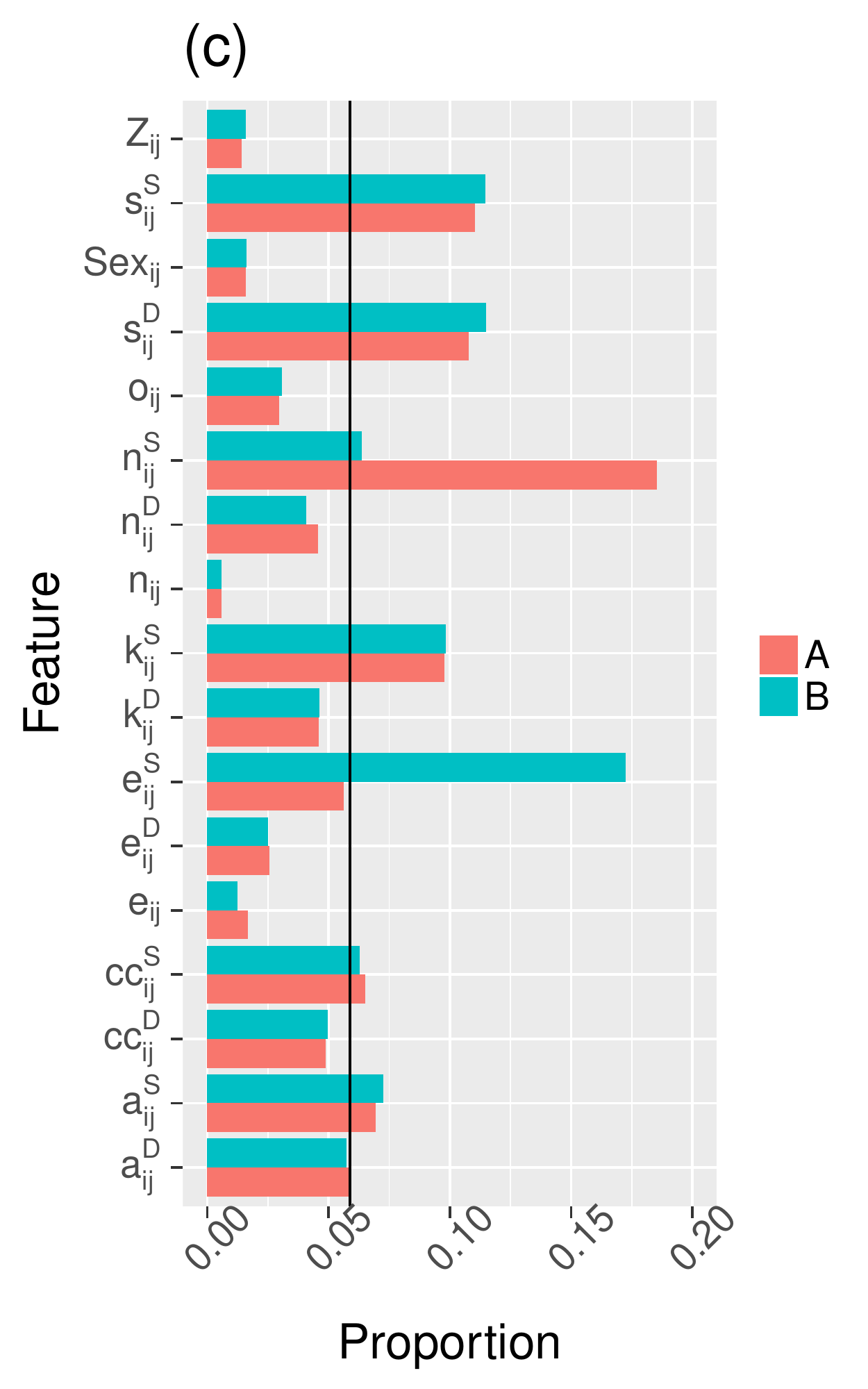}
  \caption*{}
\end{subfigure}%
\begin{subfigure}{.5\textwidth}
  \centering
  \includegraphics[width=0.7\linewidth]{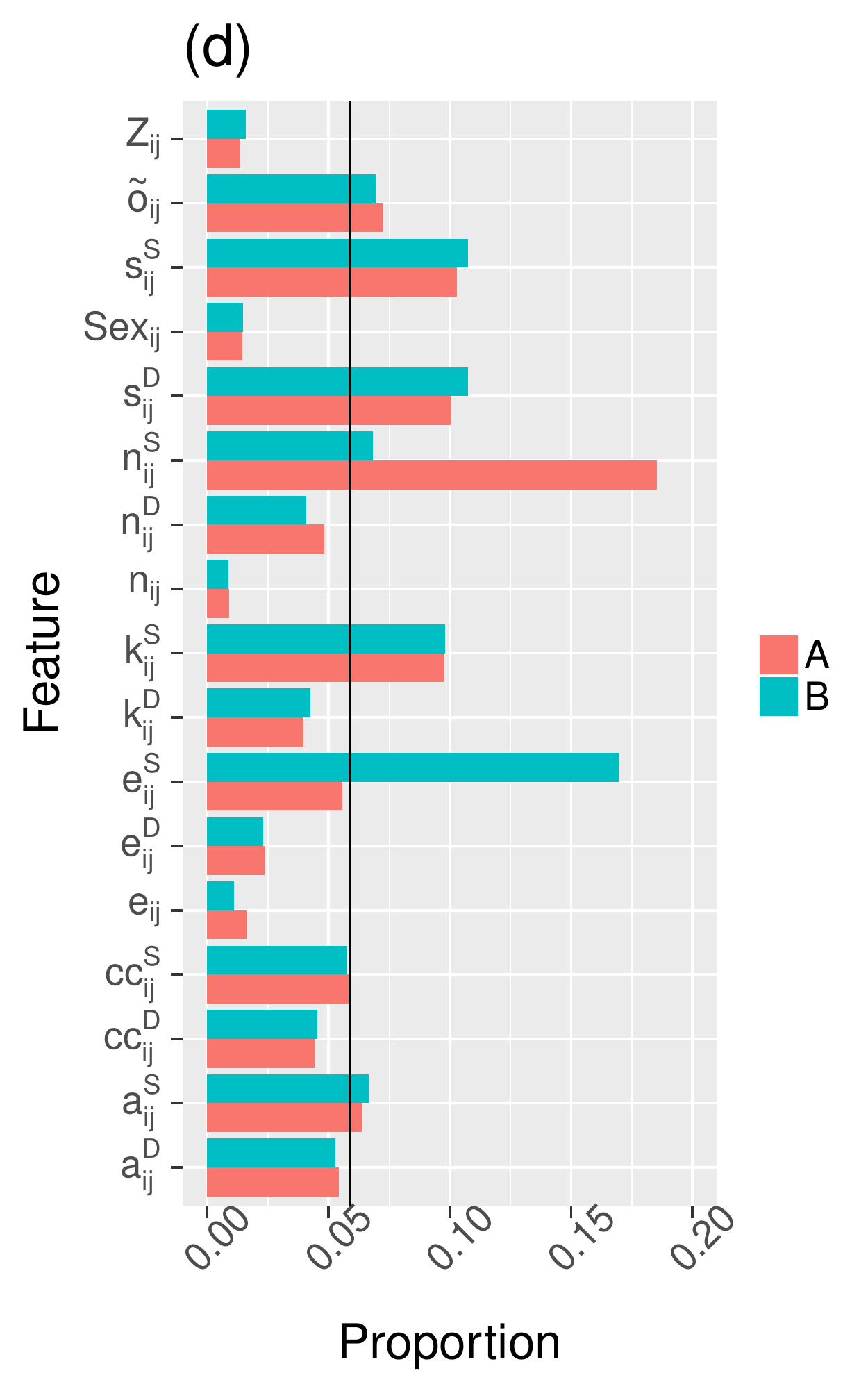}
  \caption*{}
\end{subfigure}
\caption*{\textbf{Fig. S2.} Accuracy and feature importance plots for the averaged normalized tie strength ($z_{ij}$) CDR call network. Accuracy using RF regression before (B) and after (A) imputation for all three models is shown in (a). Feature importance using RF regression before and after imputation are shown for Model 1 (b), Model 2 (c) and Model 3 (d). }
\end{figure}


\begin{figure}
\begin{subfigure}{.5\textwidth}
  \centering
  \includegraphics[width=\linewidth]{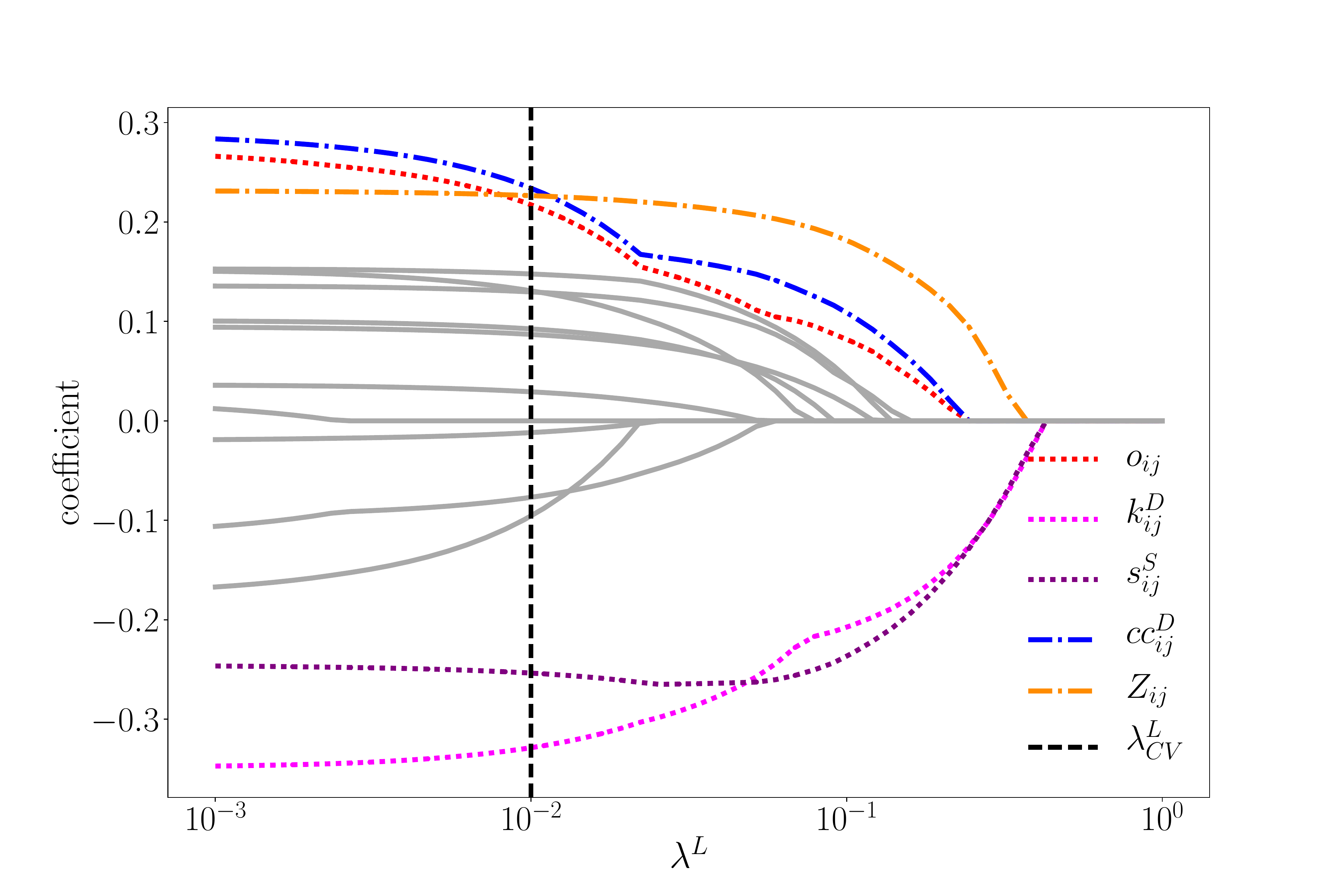}
  \caption*{(a)}
\end{subfigure}%
\begin{subfigure}{.5\textwidth}
  \centering
  \includegraphics[width=\linewidth]{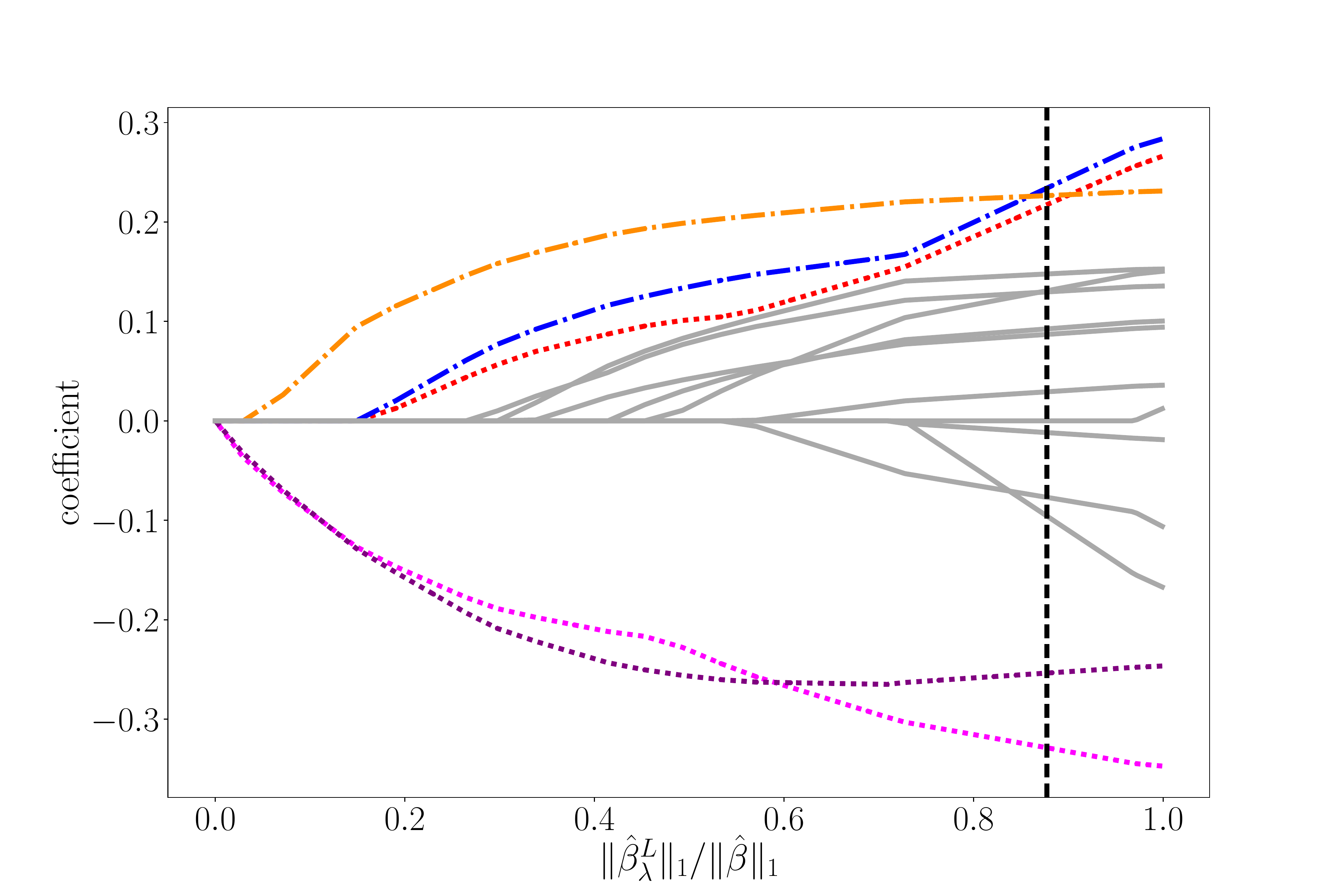}
  \caption*{(b)}
\end{subfigure}%
\\
\begin{subfigure}{.5\textwidth}
  \centering
  \includegraphics[width=\linewidth]{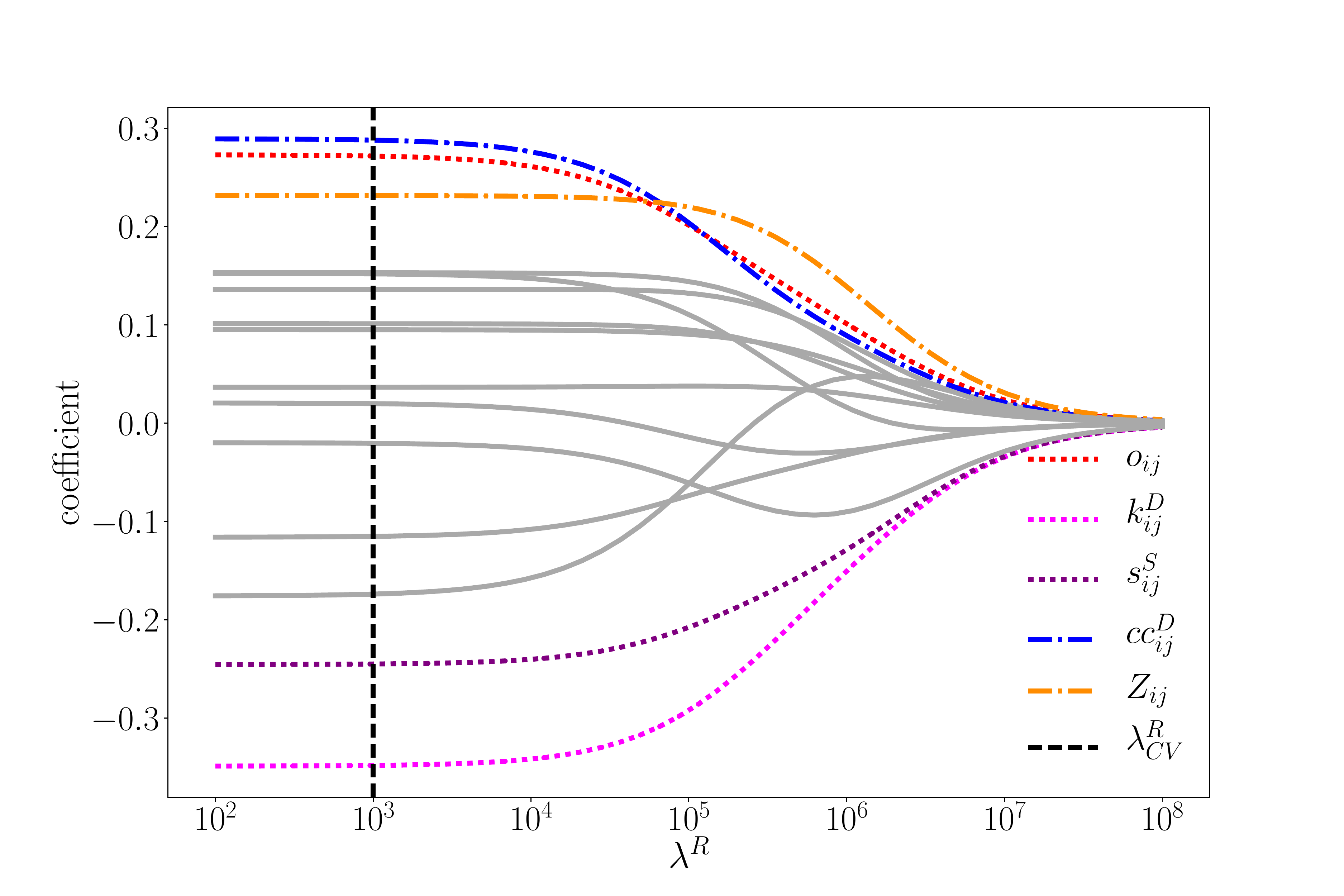}
  \caption*{(c)}
\end{subfigure}%
\begin{subfigure}{.5\textwidth}
  \centering
  \includegraphics[width=\linewidth]{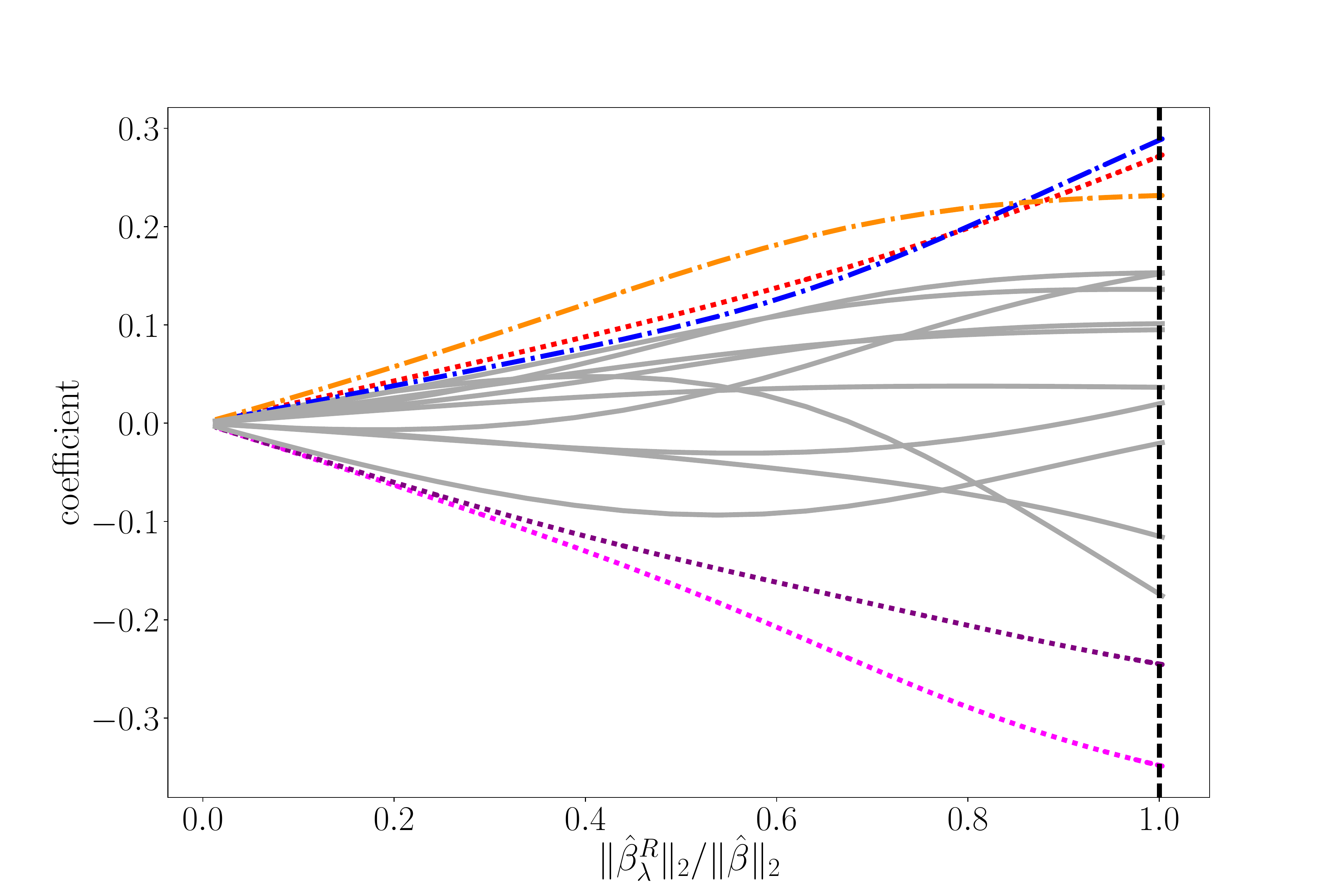}
  \caption*{(d)}
\end{subfigure}
\caption*{\textbf{Fig. S3.} The standardized LASSO coefficients as a function of $\lambda^L$ (a) and $\| \hat \beta_L \| / \| \hat \beta \|_1$ (b) using 10-fold cross validation for CDR normalized tie strength ($y_{ij}$) after imputation. Each line represents a different predictor with colored lines representing significant predictors. The dashed black line indicates the value of $\lambda^L$ chosen via cross validation and denoted as $\lambda^L_{CV}$. The standardized ridge regression coefficients as a function of $\lambda^R$ (a) and $\| \hat \beta_L \| / \| \hat \beta \|_2$ (b) using 10-fold cross validation for CDR normalized tie strength $(y_{ij})$ after imputation.The dashed black line indicates the value of $\lambda^R$ chosen via cross validation, which we denote as $\lambda^R_{CV}$.}
\end{figure}


\begin{figure}
\begin{subfigure}{.5\textwidth}
  \centering
  \includegraphics[width=\linewidth]{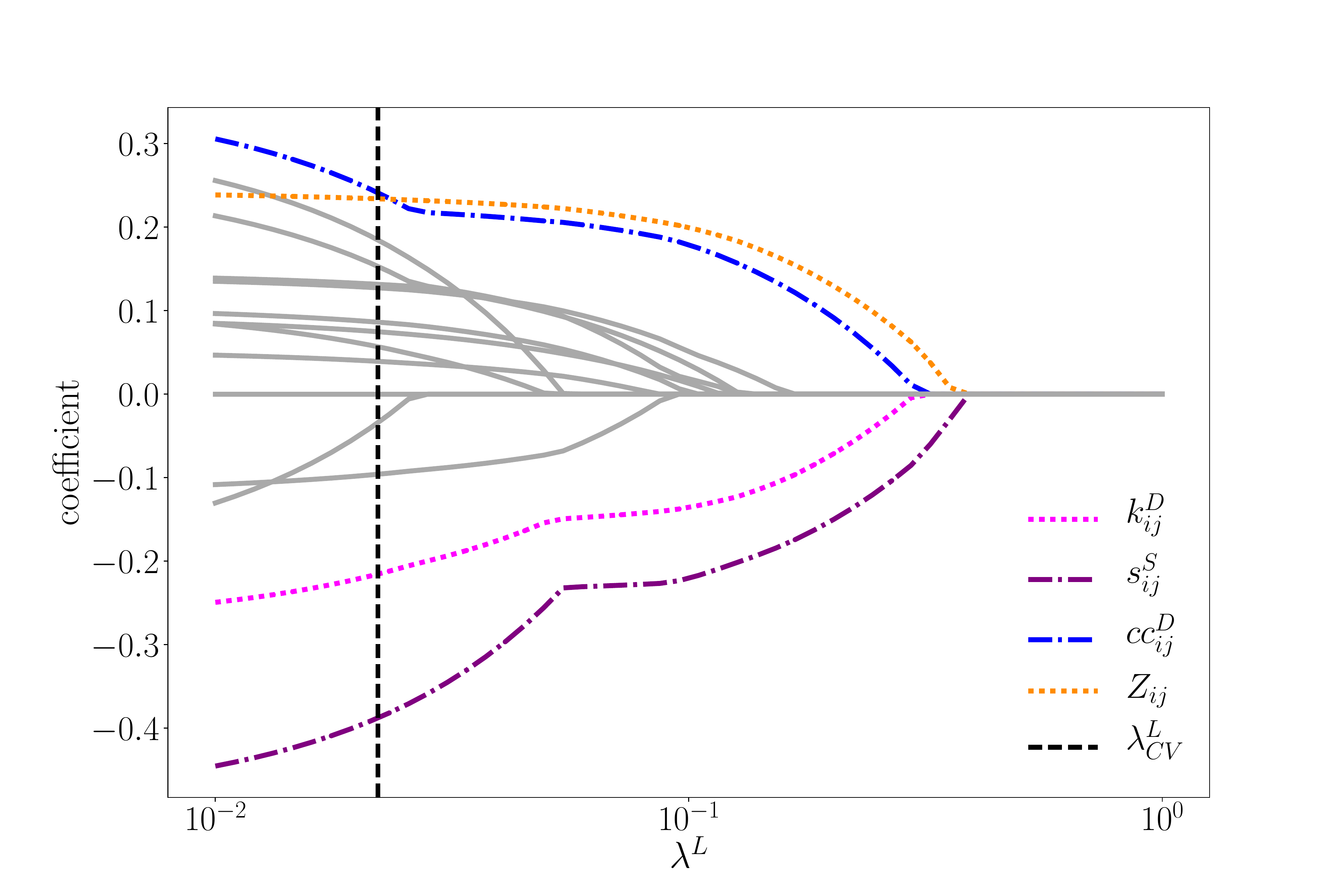}
  \caption*{(a)}
\end{subfigure}%
\begin{subfigure}{.5\textwidth}
  \centering
  \includegraphics[width=\linewidth]{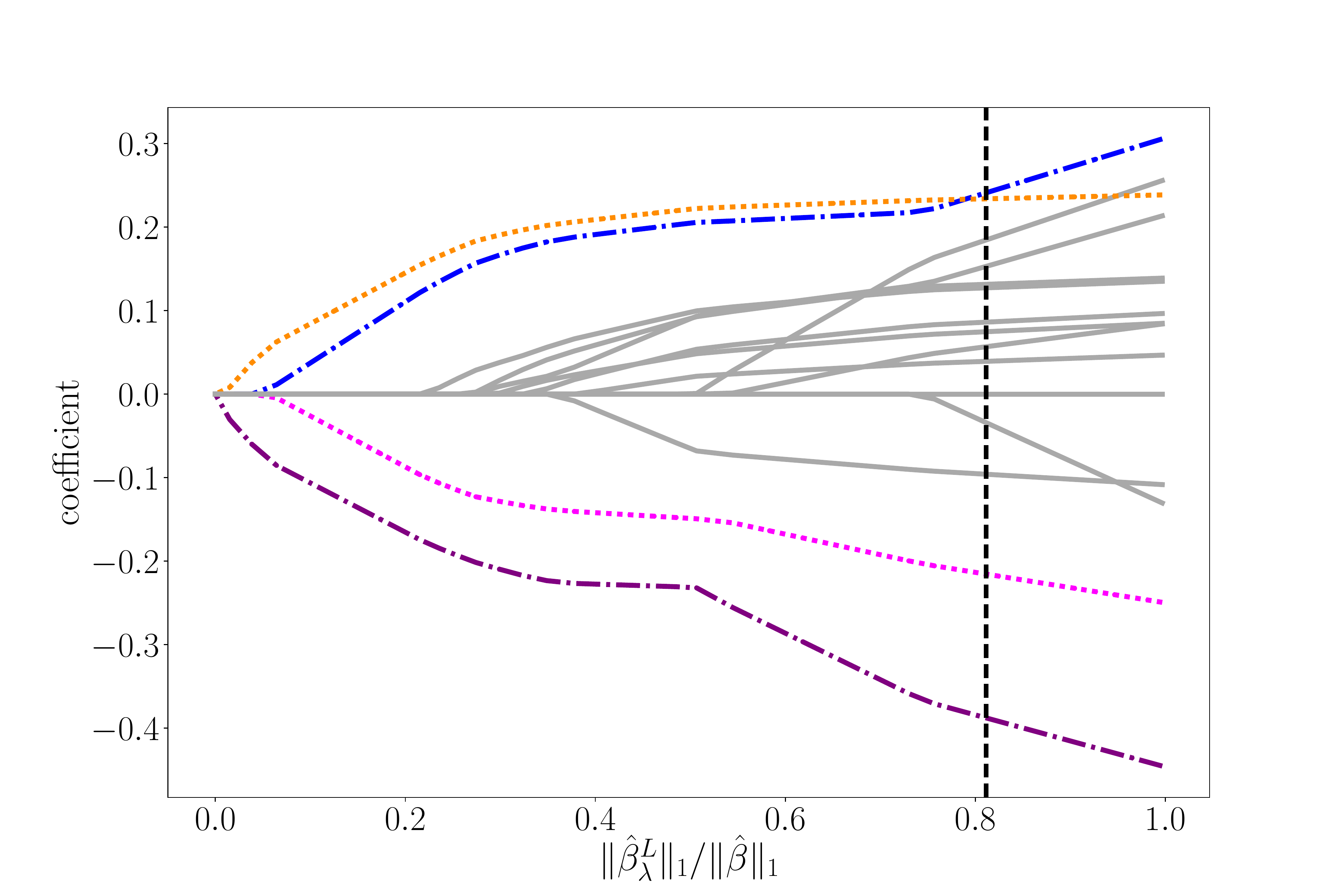}
  \caption*{(b)}
\end{subfigure}%
\\
\begin{subfigure}{.5\textwidth}
  \centering
  \includegraphics[width=\linewidth]{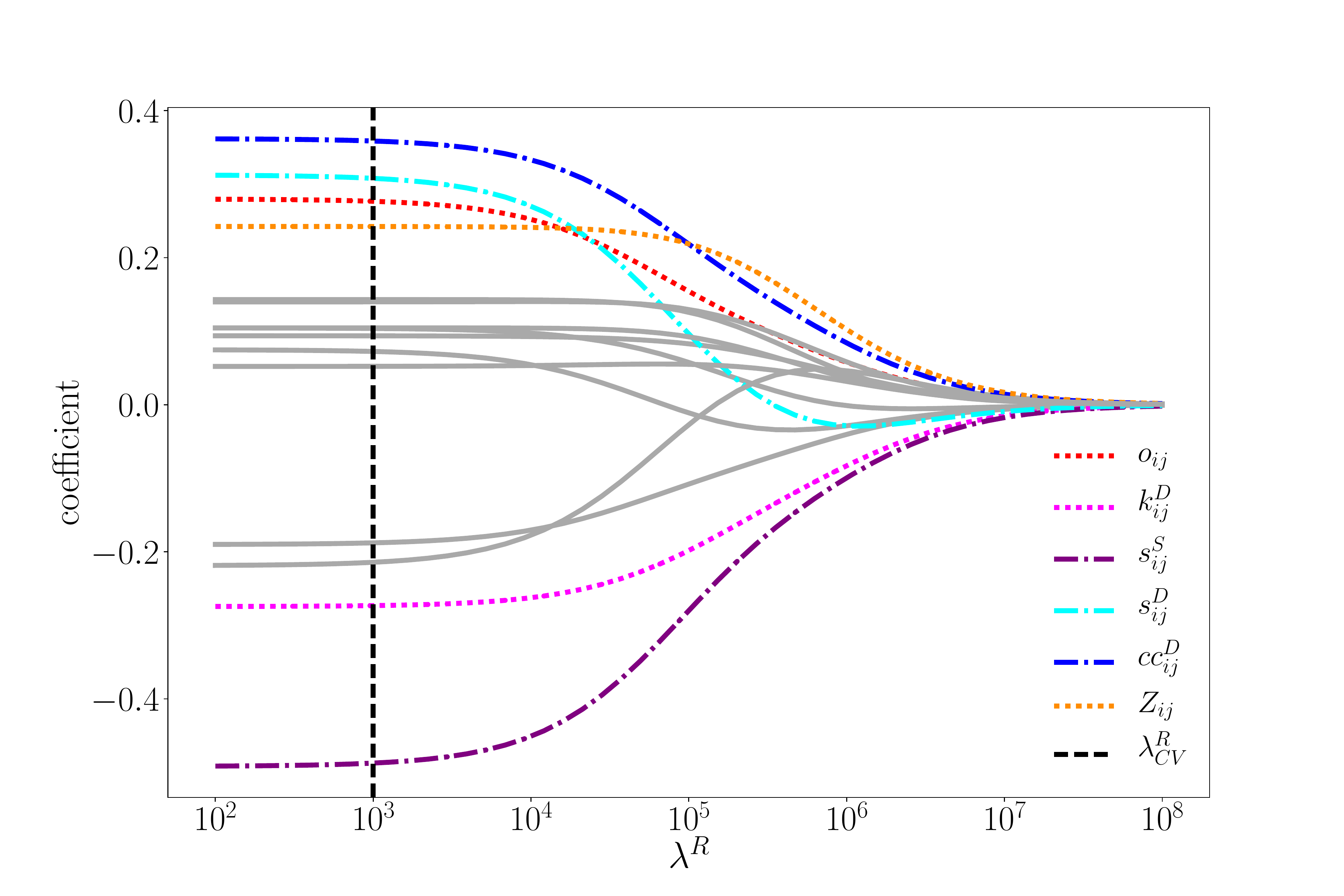}
  \caption*{(c)}
\end{subfigure}%
\begin{subfigure}{.5\textwidth}
  \centering
  \includegraphics[width= \linewidth]{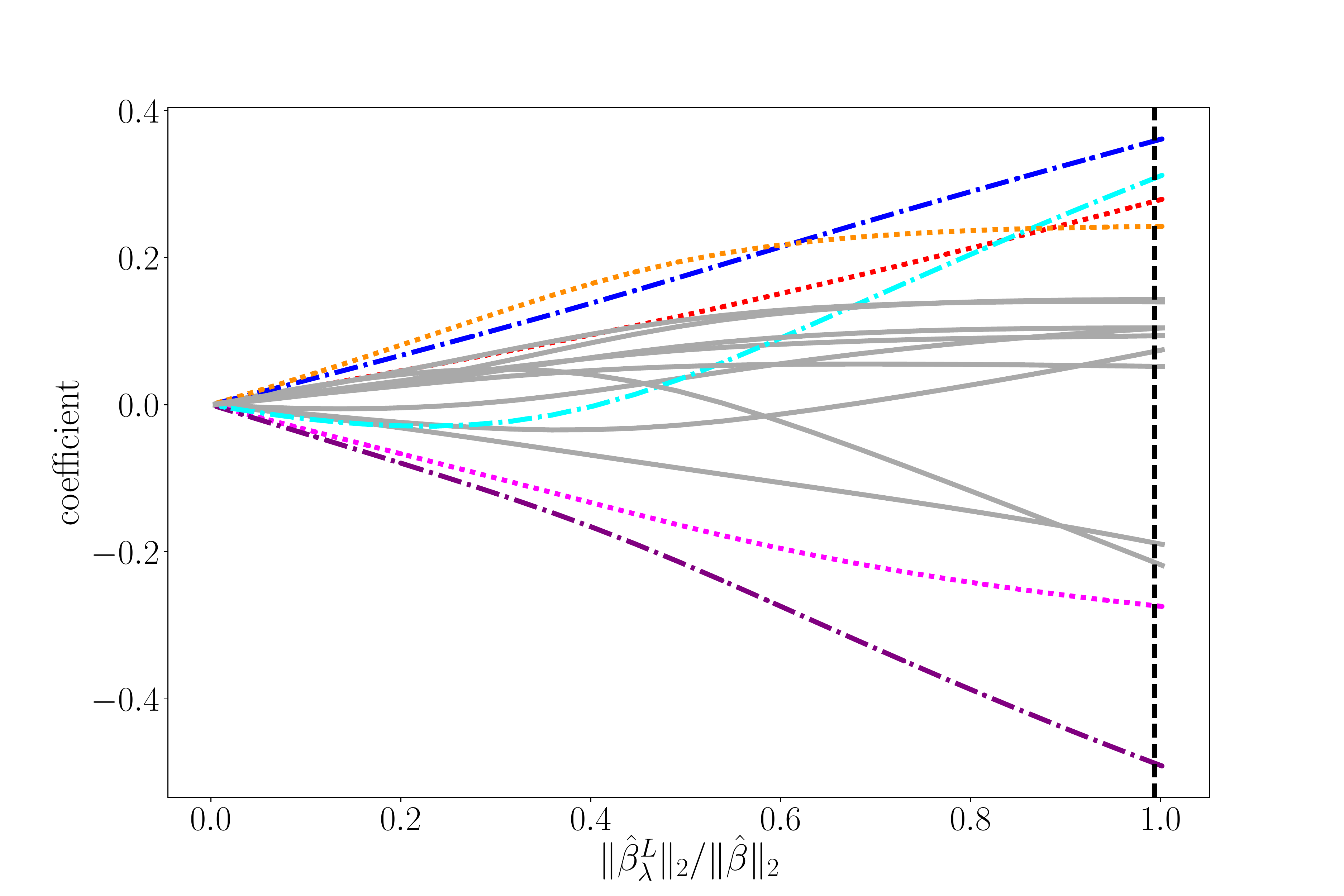}
  \caption*{(d)}
\end{subfigure}
\caption*{\textbf{Fig.S4.} The standardized LASSO coefficients as a function of $\lambda^L$ (a) and $\| \hat \beta_L \| / \| \hat \beta \|_1$ (b) using 10-fold cross validation for CDR averaged tie strength ($z_{ij}$) after imputation. Each line represents a different predictor. The colored lines represent the predictors significantly different than 0. The dashed black line indicates the value of $\lambda^L$ chosen via cross validation, which we denote as $\lambda^L_{CV}$. The standardized ridge regression coefficients as a function of $\lambda^R$ (a) and $\| \hat \beta_L \| / \| \hat \beta \|_2$ (b) using 10-fold cross validation for CDR averaged tie strength ($z_{ij}$) after imputation.The dashed black line indicates the value of $\lambda^R$ chosen via cross validation, which we denote as $\lambda^R_{CV}$.}
\end{figure}

\clearpage
\begin{table}
\centering
\label{table:CDR_coef}
\begin{tabular}{|c|c|c|l|c|c|l|c|}
\multicolumn{2}{c}{} & \multicolumn{3}{c}{Normalized Strength ($y_{ij}$)} & \multicolumn{3}{c}{Averaged Strength ($z_{ij}$)} \\
\hline
Model & Predictor & $\lambda$ & Coefficient & Adjusted $R^2$ & $\lambda$ & Coefficient & Adjusted $R^2$\\
\hline
\hline
A & $o_{ij}$ & - &  0.27 & 0.116 & - & 0.27 & 0.117\\
& $k^D_{ij}$ & & -0.35 & & & -0.35 & \\
& $s^S_{ij}$ & & -0.25 & & & -0.25 & \\
& $s^D_{ij}$ & & - & & & -0.20 & \\
& $cc^D_{ij}$ & & 0.29 & & & 0.29 & \\
& $Z_{ij}$ & & 0.23 & & & 0.23 & \\
\hline
B & $o_{ij}$ & 0.01 &  0.21 & 0.115 & 0.022 & - & 0.110\\
& $k^D_{ij}$ & & -0.33 & & & -0.21 & \\
& $s^S_{ij}$ & & -0.25 & & & -0.39 & \\
& $cc^D_{ij}$ & & 0.23 & & & 0.24 & \\
& $Z_{ij}$ & & 0.23 & & & 0.23 & \\
\hline
C & $o_{ij}$ & $10^3$ &  0.27 & 0.116 & $10^3$ & 0.28 & 0.100\\
& $k^D_{ij}$ & & -0.35 & & & -0.27 & \\
& $s^S_{ij}$ & & -0.25 & & & -0.49 & \\
& $s^D_{ij}$ & & - & & & 0.31 & \\
& $cc^D_{ij}$ & & 0.29 & & & 0.36 & \\
& $Z_{ij}$ & & 0.23 & & & 0.24 & \\
\hline
\end{tabular}
\caption*{ \textbf{Table S1.} Regression results for the CDR call network. Predictors, coefficients, shrinkage parameters $\lambda^L$ and $\lambda^R$, and adjusted $R^2$ values are reported. Model A represents OLS regression, Model B LASSO regression and Model C ridge regression.}
\end{table}


\newpage
\bibliographystyle{abbrv}
\bibliography{authors}

\end{document}